\documentclass[aps,preprint,preprintnumbers,nofootinbib]{revtex4}
\usepackage{amsmath,amsfonts,subfigure}
\usepackage{graphicx}
\usepackage{bm}
\usepackage{times}
\usepackage{graphicx}
\usepackage{cancel}
\usepackage{amsmath}
\usepackage{amssymb}
\usepackage{textcomp}
\usepackage{xcolor}
\usepackage{axodraw}

\def\st{Stueckelberg~}
\def\s1{$s_{\alpha}$}
\def\s2{$s_{\gamma}$}
\def\s3{$s_{\delta}$}
\def\c1{$c_{\alpha}$}
\def\c2{$c_{\gamma}$}
\def\c3{$c_{\delta}$}

\newcommand{\sm}{Standard Model}

\newcommand{\beqn}{\begin{eqnarray}}
\newcommand{\eeqn}{\end{eqnarray}}
\newcommand{\be}{\begin{equation}}
\newcommand{\ee}{\end{equation}}
\newcommand{\non}{\nonumber \\}

\begin{document}
\title{\Large \bf  
R-parity Conservation via the Stueckelberg Mechanism:\\  LHC and Dark Matter Signals} 
\author{ Daniel Feldman$^{1,2}$, Pavel Fileviez Perez$^{3}$, and Pran Nath$^4$}
\affiliation{$^1$ Michigan Center for Theoretical Physics, University of Michigan, Ann Arbor,
MI 48109, USA.\\
$^2$ Visitor,  CERN Theory Group, CH-1211 Geneva 23, Switzerland.\\
$^3$ Center for Cosmology and Particle Physics (CCPP), Department of Physics,  New York University, NY 10003, New York, USA.\\ 
$^4 $ Northeastern University, Department of Physics,  Boston, MA  02115, USA.}
\begin{abstract}
We investigate the connection between the conservation of R-parity in supersymmetry 
and the Stueckelberg mechanism for the mass generation of the $B-L$ vector gauge boson.
It is shown that  with universal boundary conditions 
for soft terms of sfermions in each family at the high scale and with the Stueckelberg mechanism for generating mass for the $B-L$ gauge boson 
present in the theory, electric charge conservation guarantees the conservation of R-parity in the minimal $B-L$ extended supersymmetric 
standard model. We also discuss non-minimal extensions. This includes extensions where the 
 gauge symmetries arise with an additional $U(1)_{B-L} \otimes U(1)_X$,
where $U(1)_X$ is a hidden sector gauge group. 
In this case the presence of the additional $U(1)_X$ allows for a $Z'$ gauge boson mass with $B-L$ 
interactions to lie in the sub-TeV region overcoming the multi-TeV LEP constraints. The possible 
tests of the models at colliders and in dark matter experiments are analyzed including signals
of a low mass $Z'$ resonance and  the production of 
spin zero bosons and their decays into two photons. In this model  two types of dark matter candidates emerge 
which are Majorana and Dirac particles. Predictions are made for a possible simultaneous observation of new physics events 
in dark matter experiments and at the LHC.

 \end{abstract}

\maketitle

\tableofcontents

\clearpage
\section{Introduction}
\label{sec:intro}

\hspace{.6cm} R-parity is an important symmetry in supersymmetric theories (For a review see \cite{Review}). In supergravity theories~\cite{Chamseddine:1982jx}, over most of the parameter space of models consistent with the radiative breaking of  the electroweak symmetry,  the lightest neutralino is found to be the lightest supersymmetric particle, and this, along with R-parity 
(defined as $R=(-1)^{2S + 3 (B-L)}$, where $S$, $B$ and $L$ stand for the spin, baryon number and lepton number, respectively)
and charge neutrality allows for the lightest neutralino to be a promising  candidate for cold dark matter as suggested  in~\cite{Gold}. 

The question then, is, if indeed R-parity turns out to be a conserved symmetry of nature, how does such a symmetry come about, and how one may guarantee that it is conserved. It is known that the MSSM with the inclusion of a right handed neutrino, one for each generation, has an  anomaly free $U(1)_{B-L}$ which can be gauged\footnote{A gauged $U(1)_{B-L}$ arises naturally in GUT models such as $SO(10)$ and $E_6$ and in string models.}. Of course, a $U(1)_{B-L}$ gauge boson must grow mass otherwise it would produce an undesirable long range force. 
In the analysis that follows it is shown that  a gauged  $B-L$ symmetry, where the gauge boson develops a mass through the Stueckelberg mechanism extending the Standard Model  gauge group~\cite{Kors:2004dx}
\cite{fln1,fln11} preserves R-parity, i.e., R-parity does not undergo spontaneous breaking by renormalization groups effects under the assumption of universality of soft scalar masses, 
charge conservation and in the absence of  a Fayet-Iliopoulos D-term. We will  later refer to this model as the \textit{Minimal $B-L$ Stueckelberg Extension of the MSSM.}

The fact that the minimal  gauged $B-L$ model  proposed in this work  preserves R-parity, with mass growth arising from the Stueckelberg mechanism, 
 is in  contrast to models with a gauged $B-L$ where the symmetry is broken spontaneously and thus does not necessarily preserve the R-parity invariance.
 Thus the analyses of~\cite{FP2,Barger:2008wn,pfps,Everett,FP1,FileviezPerez:2011kd} show that R-parity symmetry, even if  valid at the grand unification scale, could be broken by 
renormalization group effects~\footnote{For grand unified models where R-parity symmetry is automatic see~\cite{Martin:1992mq}. For analyses where the spontaneous breaking of $B-L$ occurs see~\cite{Khalil:2007dr,Frank:2001jp}, for  early work on the spontaneous breaking of {R-parity} see~\cite{Aulakh:1982yn,Hall,Mohapatra,MasieroValle} .
For early analyses with  R-parity and additional gauge fields see~\cite{Ibanez}.}

We will first discuss the minimal $(B-L)$ \st extension of the Standard Model and of the minimal supersymmetric Standard Model (MSSM). In these extensions the $Z'$ boson~\footnote{For recent dedicated work on heavy $Z'_{B-L}$ physics see  \cite{Accomando,BassoPruna}.} is constrained to be rather heavy, i.e., it lies in the  multi-TeV range and 
thus a direct detection may be difficult. However, this constraint is overcome in a $U(1)_{B-L}\otimes U(1)_X$ Stueckelberg extension, where $U(1)_X$ is the hidden sector gauge group. Here the Stueckelberg sector generates two extra massive vector neutral bosons, i.e., $Z'$ and $Z''$, one of which would be very narrow and could lie even in the sub-TeV region, and thus would be accessible
at the  LHC. The models with massive mediators arise generally via  mass mixing and kinetic mixing of Abelian gauge bosons~\cite{darkforce1,darkforce2,darkforceA,rev,darkforce3,darkforce4,darkforce5,darkforce6,darkforce7,darkforce8, darkforce9},~\cite{Nath:2008ch,Liu,darkforcereview} and the mixings are also the source of  the so called dark forces~\cite{darkforce1,darkforceA} - the mixings allow for a portal between the hidden (dark) sector via massive mediators~\cite{darkforce1,darkforce2,darkforceA,rev,darkforce3,darkforce4} (from which  several components of dark matter can arise) and the visible sector where the states charged under the the Standard Model reside. Specifically, the  class of models that we study here allows for two component (Majorana and Dirac) dark matter~\cite{Feldman:2010wy}.   
Such models  with  dark forces have received  considerable attention in the context of the recent cosmic anomalies~\cite{Feldman,Arkani,FLNN,Feldman:2010wy}; for recent additional works on dark sectors see e.g. \cite{Quevedo,Mam,ANelson,Mam2}.

The organization of this paper is as follows :  In sec.~({\ref{1}})  we propose a  $U(1)_{B-L}$ extension of the Standard Model via the \st mechanism.
In sec.~(\ref{2a})  the $B-L$ \st extension of MSSM is introduced. In sec.~(\ref{2b}) we outline the conditions 
for R-parity to be not spontaneously broken. 
In sec.~({\ref{3}}) we give a dedicated analysis of a  $U(1)_{B-L} \otimes U(1)_{X}$ extension of the MSSM via the \st mechanism and show that
the model naturally leads to a sharp $Z'$ prime resonance that can be seen at the LHC, and we 
analyze recent constraints from the Tevatron and the LHC.
Here we  also analyze the production and decay of new spin-0 particles.
These scalars are the real parts 
of the chiral Stueckelberg superfields, where the imaginary part are the axions which are absorbed  giving masses
to the $Z'$ and $Z''$. 
In sec.~(\ref{4}) we show that the model  allows for two component dark matter, one consisting of 
\emph{neutral  Dirac} dark matter and  the other of Majorana dark matter which produce a relic
abundance consistent with WMAP~\cite{WMAP}.
We also explore the detection possibility of dark matter with the recent
limits set by the XENON and CDMS collaborations~\cite{Aprile:2011hi,Ahmed:2009zw} which allows for direct detection constraints 
to be connected with the corresponding constraints on the $Z'$ production at colliders.  
 In sec.~(\ref{5}) we give an overview as to how models of spontaneous {R-parity} breaking can be distinguished
from the R-parity preserving $B-L$ extensions. Conclusions are given in sec.~({\ref{6}}).
 \section{ $B-L$  Stueckelberg Extension of the Standard Model \label{1}}
The $B-L$ extension of the Standard Model provides a natural framework to understand the origin 
of neutrino masses since the three families of right-handed neutrinos, needed to cancel all anomalies, 
are used to generate neutrino masses. 
We first consider a $U(1)_{B-L}$ \st extension of the Standard Model with the gauge group
\be SU(3)_C\otimes SU(2)_L\otimes U(1)_Y\otimes U(1)_{B-L} ~.\ee
The mass growth for the  $U(1)_{B-L}$ occurs via the \st mechanism
for which  the extended Lagrangian is given by 
\beqn
{\cal L} &=& {\cal L}^{B-L}_{\rm St}+{\cal L}^{B-L}_{\rm Yuk}+{\cal L}_{\rm SM}, \\
{\cal L}^{B-L}_{\rm St} &=& - \ \frac{1}{4} C_{\mu\nu} C^{\mu\nu} 
- \frac{1}{2} (M_{BL}C_{\mu} + \partial_{\mu} \sigma) (M_{BL} C^{\mu} + \partial^{\mu} \sigma), \\
{\cal L}^{B-L}_{\rm Yuk} &=&    \ Y_\nu \ \bar{l}_L \tilde{H} \nu_R  .
 \label{1.1}
\eeqn
Here ${\cal L}_{\rm SM}$ is the \sm ~Lagrangian, $l_L^T=(\nu_L,e_L)$ and $\tilde{H}= i \sigma_2 H^*$. 
As usual, the Standard Model Higgs is $H^T=(H^+, H^0)$. The above Lagrangian is invariant under 
the $B-L$  transformations 
\be 
\delta C_{\mu}= \partial_{\mu} \lambda, ~~\delta \sigma =- M_{BL} \lambda.
\ee 
Added to the above is a gauge  fixing term 
\be 
{\cal L}_{\rm gf}= -\frac{1}{2\xi} (\partial_{\mu} C^{\mu} + M_{BL} \xi \sigma)^2,
\ee 
so that the vector field becomes massive  while the $\sigma$ field decouples. 
Additionally the interaction Lagrangian 
\be 
{\cal L}_{\rm St}^{ \rm int} = g_{BL} C_{\mu}J^{\mu}_{BL}, \label{intBL}
\ee 
couples the \st field $C_{\mu}$ to the  conserved  $B-L$ vector current
$J^{\mu}_{BL}$.  We note that the $B-L$ gauge field $C_{\mu}$ has become 
massive with a mass $M_{BL}$ while maintaining the $U(1)_{B-L}$ invariance. Since $B-L$ continues to be 
a symmetry even after the mass growth of the $Z'$ its properties are rather different 
from the model where the $B-L$ gauge symmetry is spontaneously broken through the Higgs mechanism.  
We will return to this in a later section.
It is important to mention that in this theory the neutrinos are Dirac fermions since 
there is no way to generate Majorana masses for right-handed neutrinos as in the canonical 
$B-L$ model. This is a natural consequence coming from the Stueckelberg mechanism.

In the above, a kinetic mixing term  is possible leading to a generalized mass and kinetic mixings for a massive $U(1)$ 
which will then generally mix with the SM sector
\cite{darkforce1,Feldman:2007wj}
where the hypercharge vector boson $B$ mixes via both mass and kinetic mixings~\cite{darkforce1}.
One then diagonalizes the Stueckelberg mass and kinetic mixing together {\cite{Feldman:2007wj},\cite{Ringwald},\cite{Zhang},\cite{Burgess}.
A further generalization to multiple $U(1)s$   reads
\be 
\mathcal{L}^{ \rm KM }_{\rm St}  =
\frac{1}{2} \sum^{N_V}_{i,j , i \neq j}  \frac{    {\epsilon}_{ i j}} {2}  V_{i \mu \nu} {V_j}^{ \mu \nu}  
-\frac{1}{2} \sum^{N_S}_{n=1}(\partial_{\mu} \sigma_n +  \sum^{N_V}_{m=1} M_{nm} V_{\mu m})^2,
     \label{gen}
\ee 
with $N_V$ Abelian vectors and $N_S$ axions,  where $B_{\mu} = V_{\mu 1} $ and the other vector fields
correspond to either hidden
or visible gauge symmetries. Recent works with multiple additional $U(1)s$ have indeed
been discussed recently~{\cite{Feldman:2007wj,Feldman:2010wy,Heeck:2011md}, \cite{FLNN,Feldman,Dimopoulos}.}  Our analysis is restricted to non-anomalous extension of the Standard Model (for the anomalous case see e.g. \cite{coriano,anastasopoulos,fucito}).
In the analysis that follows we will assume the kinetic mixing is absent and instead investigate the pure Stueckelberg sector
in the absence of mass mixing of the hypercharge $B$ with the \st sector.   For recent works on the Stueckelberg Mechanism see e.g.~{\cite{Ahlers:2008qc,Goodsell:2009xc,PT,IB,Marsano,Quevedo,Cvetic:2011iq} and for early work in the context of strings see \cite{KR}.
\section{$ B-L $ Stueckelberg Extension of the MSSM \label{2a}}
Here we construct the minimal  $U(1)_{B-L}$ extension of the MSSM using the Stueckelberg Mechanism.
The supersymmetric extension of Eq. (\ref{1.1}) is
\be 
{ \cal L}_{\rm St} = 
 (M_{BL} C +  S_{\rm st} + {\bar S}_{\rm st})^2 |_{\theta^2 \bar \theta^2}\ ,
\label{mass}
\ee 
where $C=(C_\mu,\lambda_C,D_C)$ is the gauge vector multiplet for $U(1)_{B-L}$,
and the Stueckelberg multiplet is $S_{\rm st} = (\rho + i \sigma , \psi_{\rm st}, F_S)$  where $\rho$ is a scalar while $\sigma$ is the axionic
pseudo-scalar. The supersymmetrized gauge transformations under the $U(1)_{B-L}$ are
\beqn
\delta_{BL} C = \zeta_{BL} + \bar\zeta_{BL}  \ , \quad
\delta_{BL} S_{\rm st} = - M_{BL} \zeta_{BL}\ ,
\eeqn
where $\zeta$ is an infinitesimal chiral superfield.
Next we couple the chiral matter fields $\Phi_i$  consisting of quarks, leptons and Higgs fields
of MSSM.
These couplings are given by
\beqn
{\cal L}_{\rm matter} ~=~  
 \bar \Phi_m e^{ 2g_{BL} Q_{BL} C} \Phi_m |_{\theta^2 \bar \theta^2}\ 
\label{matter}
\eeqn
where $Q_{BL}\equiv B-L$ and the sum  is implicit over the chiral multiplets $m$ and the interaction
term of Eq. (\ref{intBL}) couples the $B-L$ vector field to fermions.
We focus on the bosonic part of the extended Lagrangian which is given by
\beqn \label{stmssm}
{\cal L}_{\rm spin[0,1]} &=&
-\frac14 C_{\mu\nu}C^{\mu\nu}-\frac{1}{2} M^2_{BL} C_{\mu}^{2} 
-\frac{1}{2} (\partial_\mu \rho)^2 - \frac12  M^2_{BL}  \rho^2\nonumber\\
&&- |D_\mu \tilde f_i|^2 
-    g_{BL} M_{BL} \ \rho  \  {\tilde f}_i^\dagger  Q_{
BL} \tilde f_i  
-\frac12 \Big[ \sum_i  {\tilde f}_i^\dagger g_{BL} Q_{
BL} \tilde f_i \Big]^2 \ .
\eeqn
The superpotential of the $B-L$ extended theory is simply
\beqn
{\cal W}= \mu H_u H_d + \sum_{\rm gen} [Y_u Q H_u u^c + Y_d Q H_d d^c + Y_e L H_d e^c +  Y_{\nu} L H_u \nu^c].
\label{w1}
\eeqn
Aside from the term $Y_\nu \ L H_u \nu^c$ Eq.(\ref{w1}) is the superpotential of MSSM but without 
the terms that violate R-parity. 
\section{R-parity Conservation \label{2b}}
As pointed out earlier, while the \st mechanism gives mass to the $B-L$ gauge boson, 
the Lagrangian of the theory, after the mass growth, still has a  $B-L$ symmetry and hence a 
conservation of R-parity $(R=(-1)^{2S + 3 (B-L)}=(-1)^{2S} M$. Here $M$ denotes matter parity, 
which is $+1$ for Higgs and gauge superfields, and $-1$ for all matter chiral superfields). This conservation of {R-parity} 
in the minimal $B-L$ \st  extensions is in contrast to models where the $B-L$ gauge symmetry 
is broken by a Higgs mechanism and where in general the mass growth of the $B-L$ gauge boson could break the  $B-L$ symmetry and thus R-parity invariance  is also lost. 
For example, for the model of Eq.~(\ref{w1}),  a VEV growth  for the scalar field in the $\nu^c_l$ multiplet will break $B-L$ invariance and generate a mass 
for the $B-L$ gauge boson. However,  a VEV growth for  $\tilde \nu^c_l$ also violates R-parity invariance which then removes the neutralino as a possible candidate for dark matter. 
Specifically, for example, in Eq.~(\ref{w1}) the VEV growth  of $\tilde \nu^c_l$ generates the term $LH_u$ which breaks R-parity. However, in 
 the minimal $B-L$ \st extension of MSSM even after the mass growth of the $B-L$ gauge boson R-parity is maintained and the R-parity
violating  interactions such as  $LH_u$, $LLe^c$, $QLd^c$,  $u^c d^c d^c$  are all forbidden in the superpotential.  

\subsection{\it Scalar Potential and  R-Parity Conservation}
We wish to show here that  with a \st mechanism for mass generation the $B-L$  symmetry not only remains 
unbroken at the tree level but further that this invariance is not violated by radiative breaking  in the minimal model. 
We give now the deduction of  this result which is rather straightforward. We exhibit below the potential including just one 
generation of leptonic scalar fields in the model consisting of  $\rho,  \tilde \nu, \tilde e,  \tilde e^c,  \tilde \nu^c$ 
(An extension to 3 generations is trivial). 
Assuming charge conservation so that $\langle \tilde e\rangle=0= \langle \tilde e^c \rangle,~\rm etc.,$
and  including soft  breaking,  the potential that involves $\rho$, $\tilde \nu$ and $\tilde \nu^c$ fields is
 \beqn
 V_{{\rm St}-BL} &=& \frac{1}{2} \left(M_{BL}^2 + m_{\rho}^2 \right) \rho^2  
+ 
 M_{\tilde\nu}^2  \tilde \nu^{\dag}\tilde \nu   
 + 
 M_{\tilde\nu^c}^2   \tilde{\nu}^{c\dag} \tilde{\nu}^c     
 \nonumber\\
& +& 
\frac{g^2_{BL}}{2} \left(
 \tilde\nu^{c\dag} \tilde \nu^c -\tilde \nu^{\dag} \tilde \nu \right)^2 
  +  g_{BL} \ M_{BL} \  \rho \ \left(
 \tilde\nu^{c\dag} \tilde \nu^c
 -\tilde \nu^{\dag} \tilde \nu 
 \right), \nonumber \\
 &+& |Y_{\nu}|^2(|H_u^0\tilde\nu^c|^2 + |\tilde\nu H^0_u|^2 + |\tilde\nu\tilde\nu^c|^2) \nonumber\\
 &-& \frac{1}{4}(g^2+ g^{\prime 2}) (|H_u^0|^2-|H_d^0|^2)|\tilde \nu|^2+ \frac{1}{8}(g^2+g^{\prime 2})|\tilde\nu|^4\nonumber\\
  &+&(-\mu^* Y_{\nu} H_d^{0*}\tilde\nu\tilde\nu^c+ A_{\nu} Y_{\nu} \tilde\nu\tilde\nu^c H_u^0 + h.c.)   
 \label{V1}
 \eeqn
 where we have used $Q_{BL}(e)=Q_{BL}({\nu})= -1$
and where $m_{\rho}$, $M_{\tilde \nu}$,  and 
$M_{\tilde \nu^c}$ are soft masses. 
The relevant part of the potential is then
 \be
 V={\sum}_{\rm gen}V_{{\rm St}-BL} + V_{\rm MSSM},
 \ee
and where as is familiar
 \beqn
V_{\rm MSSM} \!&=&\!
(|\mu|^2 + m^2_{H_u}) |H_u^0|^2 + (|\mu|^2 + m^2_{H_d}) |H_d^0|^2
- (B \mu\, H_u^0 H_d^0 + {\rm h.c.})
\nonumber \\ && 
+ {1\over 8} (g^2 + g^{\prime 2}) ( |H_u^0|^2 - |H_d^0|^2 )^2 .
\eeqn
We begin with universal boundary conditions for the RGEs.
We note  that the RG evolution for $M_{\tilde e}$ and $M_{\tilde \nu}$ are identical since 
$SU(2)_L \otimes  U(1)_Y$ symmetry is unbroken down to electroweak scale.  If  
$M_{\tilde e}^2$  turned tachyonic it would lead to VEV formation for the field 
$\tilde e$ violating charge conservation  and thus we disallow this possibility. 
Since  $\tilde\nu$ and $\tilde e$  lie in the same $SU(2)_L$ multiplet the 
same holds for the $\tilde\nu$ field, i.e., it too does not develop a VEV.
This can be seen from the one loop RG sum rule connecting the sneutrino  $\tilde\nu$ mass and the 
selectron mass
  \be
    M_{\tilde\nu}^2 - M_{\tilde e}^2 =  \cos(2\beta) M_W^2 +\delta^2_{\nu,e} ,   
    \label{a8}
    \ee 
  where $\delta^2_{\nu,e}$ is difference of the mass squares of the  fermions
  (and is essentially negligible compared to $W$ mass term
    the largest of which occurs for $e \to \tau$ which is still negligible).
    Thus the right hand side of Eq.(\ref{a8}) is  positive definite for any range of $\tan\beta$ in the perturbative domain
 in the RG analysis. As a consequence,  if the mass square
    of $\tilde e$ does not turn tachyonic, this also holds for the mass square of $\tilde \nu$
    and $\langle \tilde \nu \rangle=0$.  
Thus with  $\left<\tilde e\right>=0= \left<\tilde \nu\right> = \left<\tilde e^c\right>$ ,
and integrating on the $\rho$ field, we get the following potential for $\tilde \nu^c$
\beqn
V_{\tilde{\nu}^c}= M_{\tilde\nu^c}^2  \tilde \nu^{c\dag}\tilde \nu^c   + 
\frac{ g^2_{BL} m_{\rho}^2}  { 2 (M_{BL}^2 + m_{\rho}^2)} (\tilde \nu^{c\dag} \tilde\nu^c)^2
+ |Y_{\nu}|^2 |H_u^0\tilde\nu^c|^2.
\eeqn
The last term above is negligible in size compared to the other terms since it involves
the Yukawa $Y_{\nu}$. Thus the coupling between this sector and the MSSM sector via the
$H^0_u$ field is negligible.
Now in the  RG analysis there are no beta functions to turn $M_{\tilde\nu^c}^2$ negative 
and the quartic term is positive definite so the potential is bounded  from below. 
Consequently the potential cannot support spontaneous breaking to generate a VEV for the field
$\tilde\nu^c$ and thus $\left<\tilde \nu^c\right>=0$.
 Further, the extrema equation for $\rho$ gives 
\beqn
\left< \rho \right> = -\frac{g_{BL} M_{BL}}{M_{BL}^2 \ + \  m_{\rho}^2}  
\left<
 \tilde {\nu^c}^{\dagger} \tilde {\nu^c}  - \tilde \nu^{\dagger}  \tilde \nu  \right>  = 0,
\eeqn
and since $\left<\tilde \nu\right>=0=\left<\tilde \nu^c\right>$,  
one also has $\left<\rho\right>=0$. Thus there is  no spontaneous symmetry breaking 
in the system and the  $B-L$ and consequently an R-parity is preserved.
  We add that the situation here
is rather different
from the Stueckelberg extensions introduced in \cite{Kors:2004dx,fln1,fln11} where $\rho$ receives a non-vanishing VEV.
In \cite{Kors:2004dx,fln1,fln11}, a non-vanishing VEV for $\rho$ would arise due the \st sector mixing with the $U(1)_Y$ sector of MSSM. 
 In contrast in the minimal $B-L$ extension analyzed here  there is no mixing with the $U(1)_Y$ sector, 
and thus there is no VEV growth for $\rho$.  
 Thus the entire mass growth in the $U(1)_{B-L}$ sector occurs via the \st mechanism.
If we include a Fayet-Iliopoulos D term~\cite{Fayet:1974jb} then effectively 
  the potential for $\tilde \nu^c$ is replaced by
\beqn
V_{\nu^c}= M_{\tilde{\nu}^c}^2   \tilde \nu^{c\dag} \tilde \nu^c   + 
\frac{ g^2_{BL} m_{\rho}^2}  { 2 (M^2_{BL} + m_{\rho}^2)} (\tilde \nu^{c\dag} \tilde \nu^c + \xi)^2+ |Y_{\nu}|^2 |H_u^0\tilde\nu^c|^2.
\eeqn
For the case when $\xi$ is negative a VEV growth for $\tilde \nu^c$ is possible and  R-parity can be broken spontaneously.  
While an FI~D-term naturally arises when the $U(1)$ is anomalous the inclusion of an
FI term for a non-anomalous $U(1)$, which is the case we discuss, is superfluous, and we exclude it 
 from the minimal model. Therefore, \textit{it is apparent that R-parity is always conserved 
within the minimal Stueckelberg $ B-L $ extension of the MSSM. }

The analysis above follows with (minimal) universal boundary conditions on
the soft scalar masses. However, since the nature of physics at the Planck scale is still largely 
unknown one should consider non-universalities  as well. In this case one will have 
additional contribution to the mass squares of scalar masses~\cite{Martin:1993zk,Nath:1997qm}.
The analysis of ~\cite{Ambroso} considers a contribution to 
$M_{\tilde \nu^c}^2$ arising from  $Tr(Q_{BL} m^2)$ 
with
\beqn
 S_{BL} \equiv Tr(Q_{BL} m^2)
= 2(M_{\tilde Q}^2  - M_{\tilde  L}^2) + (M_{\tilde e^c}^2 -M_{\tilde  d^c}^2)
+ (M_{\tilde \nu^c}^2- M_{\tilde u^c}^2),
\label{X}
\eeqn
under the constraint $Tr(Ym^2)=0$, where
\beqn
  S_{Y} \equiv Tr(Ym^2) = M_{H_2}^2 -M_{H_1}^2 + \sum_{\rm gen} (M_{\tilde Q}^2  - 2 M_{\tilde u^c}^2 
+ M_{\tilde  d^c}^2 - M_{\tilde L}^2 + M_{\tilde e^c}^2).
\eeqn
With the universal boundary conditions for only each family one has $S_{BL}=0$. 
This can be achieved in {minimal supergravity models}  where all scalars have the same soft mass term, 
or in $SO(10)$ or $E_6$ scenarios where the boundary conditions tell us that all sfermions of one family should have the same soft mass term.  
However, with non-universal boundary
conditions one will have in general  $S_{BL}\neq 0$.  With inclusion of
$S_{BL}$ one could in principle turn $M_{\tilde \nu^c}^2$ negative. Such a situation is achieved
  with inclusion of specific constraints in the  analysis of~\cite{Ambroso}.
However, such constraints are not generic and the positivity 
$M_{\tilde \nu^c}^2$ 
may still be
broadly valid even with inclusion of non-universalities of soft parameters. 

Now there are stringent bounds on an extra $B-L$ type gauge boson. One finds ~\cite{Carena:2004xs} 
  \beqn
M_{Z^{'}}/g_{BL} > M_{BL} \sim 6 ~{\rm TeV},
\label{c1}
\eeqn
which implies that for $g_{BL}\sim 1$ the $B-L$ type $Z'$ boson lies in the several TeV region.
With a $Z'$ of this mass scale, detection at LHC-7 may be difficult, both because of energy  considerations and luminosity. 
  Further, with the constraint as given by Eq.(\ref{c1})
some of the other phenomenological implications of the model associated with the spin 0 and spin $\frac{1}{2}$ 
sectors will also be difficult to test.
 In what follows, we uncover a model which maintains the strict R-parity invariance 
of the minimal \st $B-L$ extensions, even after mass growth of the $B-L$ gauge bosons, but
with testable implications that are far more rich.

\section{{\boldmath $U(1)_{B-L} \otimes U(1)_X $}  Stueckelberg Model \label{3}}
As indicated in the last section, the $Z'$ boson of  the minimal $B-L$ model may be difficult to detect because 
of its heavy mass. We consider now  an extension of the model of the previous section which 
overcomes this constraint and produces a $Z'$ which is much  lighter but still has $B-L$ interactions
with matter. This extension   includes 
a hidden  sector $U(1)_X$ which is anomaly free but allows for a mixing between the visible and the hidden sectors.
The extended gauge group reads:
\be SU(3)_C\otimes SU(2)_L\otimes U(1)_Y\otimes U(1)_{B-L}  \otimes U(1)_{X} ~.\ee
Thus we  have  \st  mass growth in the Abelian sector via the interaction 
\beqn
{ \cal L}_{\rm St} = \int d^2 \theta d^2\bar\theta \ 
[(M_1 X + M_2' C+  S +\bar S)^2\non
+ (M_1' X + M_2 C+  S'+ \bar S' )^2]\ ,
\label{mass}
\eeqn
where the model is invariant under the extended gauge transformations 
\beqn
\delta_X (X,S,S',C) &=&(\epsilon_X + \bar \epsilon_X , -M_1 \epsilon_X,-M_1' \epsilon_X,0) \non
\delta_{BL} (C,S,S',X) &=&(\epsilon_{BL} +\bar \epsilon_{BL},-M_2' \epsilon_{BL},-M_2 \epsilon_{BL},0)
\eeqn
where $\epsilon_{X,BL}$ are infinitesimal chiral superfields.
One can compute the mass matrix for the $U(1)_X$ and the $U(1)_{B-L}$ gauge vector bosons 
by going  to the unitary gauge which in the basis $X_{\mu}, C_{\mu}$ gives  
\beqn
 \label{neutrmass}
M^{2}_{\rm [spin ~1]} = 
\left[
\begin{matrix}
 M_1^2  +  M^{\prime 2}_1   & M_1M_2'+ M_1' M_2 \\
M_1M_2'+ M_1' M_2 & M_2^{2} + M^{\prime 2}_2
\end{matrix}
\right]  \ .
\label{vectormass}
\eeqn
Here $M_1', M_2'$ are the mixing parameters and in the limit that $M_1', M_2' \to 0$ we have
that the masses of the $X_{\mu}, C_{\mu}$  bosons are $M_1, M_2$.  The diagonalization gives  us two 
massive vector 
bosons which we may call $Z', Z''$ where 
\beqn
X_{\mu}&=& \cos \theta_{BL} Z'_{\mu}  + \sin \theta_{BL}   Z_{\mu}'',\non
C_{\mu}&=& -\sin\theta_{BL}  Z'_{\mu} + \cos \theta_{BL} Z_{\mu}''.
\label{z1}
\eeqn
We consider now the case 
of small mixing, i.e., $M_1', M_2' \ll M_1, M_2$ which implies 
 $\tan \theta_{BL} \ll 1$.  For small mixings 
 the $Z'$ boson lies mostly in the hidden sector
with a small component proportional to $\tan\theta_{BL}$ in the $B-L$ sector while the opposite holds
for $Z''$. Here $Z''$ lies mostly in the $B-L$ sector with a small component proportional to $\tan\theta_{BL}$
in the hidden sector.  
 \begin{table}[t!]
\begin{center}
\begin{tabular}{cccc|cc}
\hline
 $f\bar f$                  &   $\Gamma(Z'\to f\bar f)/\alpha_{BL} M_{Z'}$  &  $\Gamma(Z''\to f\bar f)/\alpha_{BL} M_{Z''}$ \\
 \hline
 $\ell^+_i  \ell^-_i$        &  $\sin^2\theta_{BL}/{3} $ &  $\cos^2\theta_{BL}/{3}$ \\
  $\nu_{\ell_i}  \nu_{\ell_i}$        &  $\sin^2\theta_{BL}/{3} $ &  $\cos^2\theta_{BL}/{3}$ \\
$q\bar q (q\neq t)$ &   $ f_s \sin^2\theta_{BL}/{9}$ &  $f_s \cos^2\theta_{BL}/9 $ \\ 
 $t\bar t$                   &   $f_s  f_{t,Z'} \sin^2\theta_{BL}/9 $   &   $f_s   f_{t,Z''} \cos^2\theta_{BL}/{9} $  
\end{tabular}
\caption{ 
The decay widths of the $Z'$ and of the $Z''$ bosons into leptons and into quarks in the $U(1)_X\otimes U(1)_{B-L}$ \st model
where $\alpha_{BL}\equiv g^2_{BL}/4\pi$ and $ f_s =(1+\frac{\alpha_s}{\pi})$ and for $V =Z',Z''$,  one has $ f_{t,V} = (1+2\frac{m_t^2}{M_{V}^2}) (1-\frac{4m_t^2}{M_{V}^2})$. } \label{Tab1}
\end{center}
\end{table} 

Since $X_{\mu}$ lies in the hidden sector and has no couplings to the visible
sector matter, the only couplings of $Z', Z''$ to the visible sector arises because of the couplings of 
$C_{\mu}$ to the visible sector matter.  
Using the  couplings of $C_{\mu}$  one finds the
couplings of $Z'$ and $Z''$ to the fermions $(f_i)$ to be of the form 
\beqn
{\mathcal L }_{Z',Z''} = (\bar f_{i} \gamma^\mu g_{BL} Q_{BL} f_{i})  [-\sin \theta_{BL} Z_\mu'   
 +  \cos \theta_{BL} Z_\mu''].
\label{zz}
\eeqn
In the context of Eq.(\ref{zz}) the constraint of Eq.(\ref{c1})  gives two separate conditions, i.e., 
  \beqn
M_{Z^{'}}/g_{BL} > \sin\theta_{BL}\times (6~{\rm TeV}), \non
M_{Z^{''}}/g_{BL} >  \cos\theta_{BL} \times   (6~{\rm TeV}).
\label{c2}
\eeqn
It is clear that the constraint on the $Z'$ is now considerably weakened relative to the constraint of 
Eq.(\ref{c1}) if the mixing angle $\theta_{BL}$ is 
small and one can have 
\be
M_{Z'}  \ll  {\rm 1 ~TeV},  ~~~~ {\rm \st} ~~U(1)_{B-L} \otimes U(1)_{X}.
\ee
However,  $Z''$ is still heavy since $\cos\theta_{BL}\sim 1$ for small $\theta_{BL}$. 
\begin{figure*}[t!]\centering
\includegraphics[height=6.6cm]{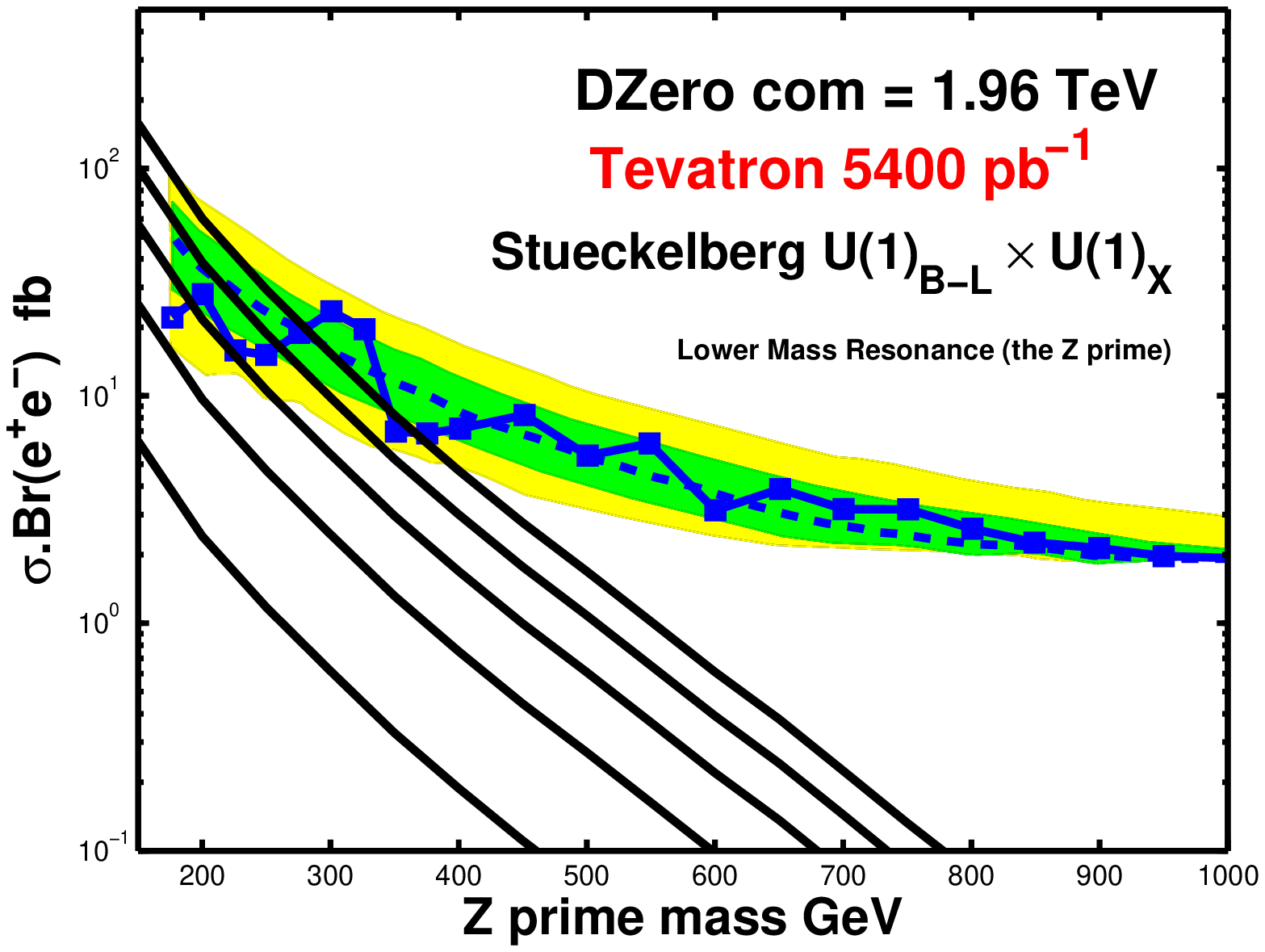}
\includegraphics[height=6.6cm]{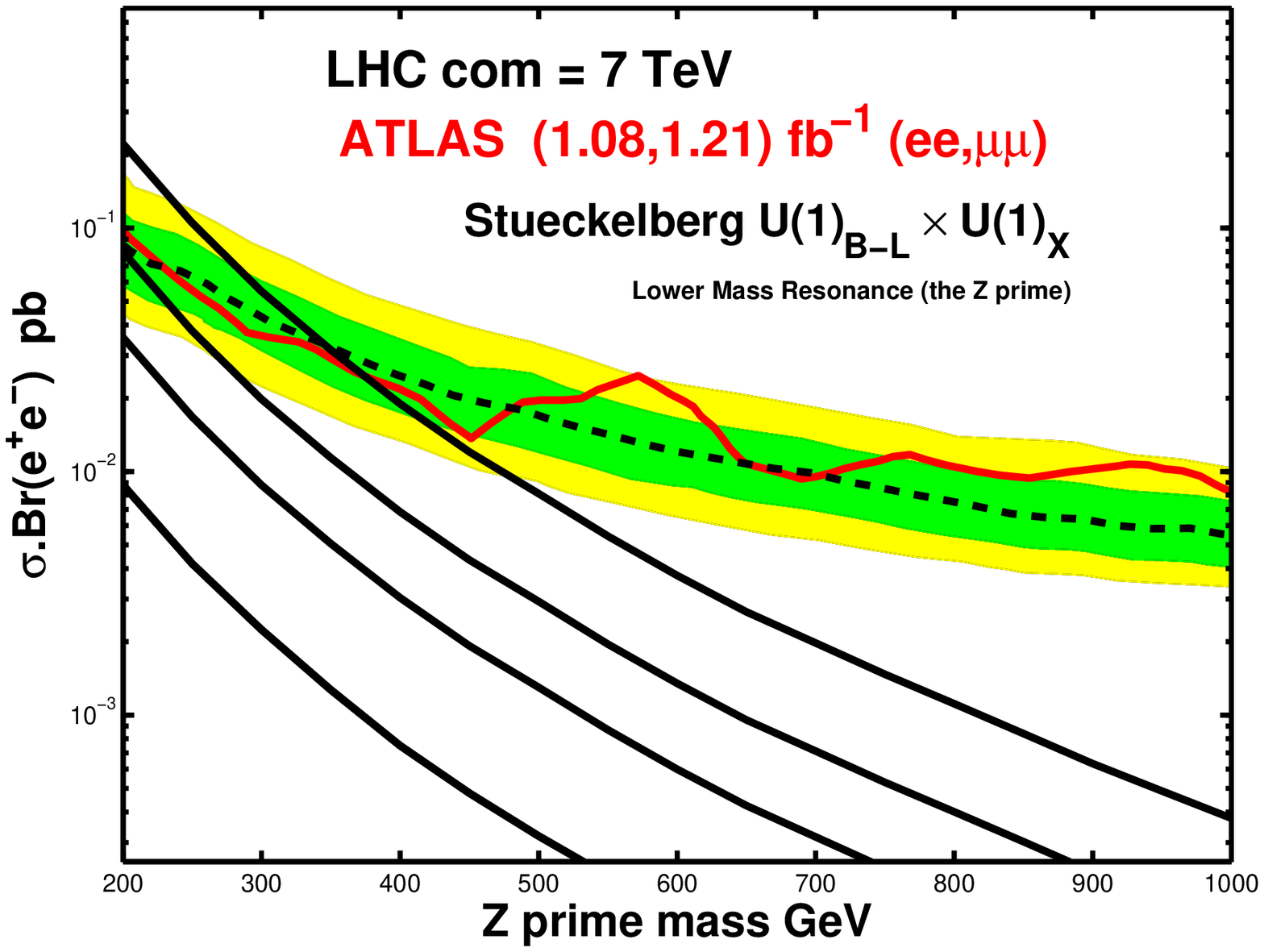}
\caption{
\label{zprime1}
Upper panel: An exhibition of $\sigma(pp\to Z')\cdot Br(Z'\to e^+e^-)$ vs the  mass of the $Z'$ resonance 
  in the \st $U(1)_{B-L}\otimes U(1)_X$ extension of MSSM at the Tevatron. Here $g_{BL}= 0.35$ and $\sin\theta_{BL}$ takes on the
  values $0.01,0.05$ from the bottom to the top curves in the plot.    
  The analysis assumes that the
  $Z'$ decay into the hidden sector is suppressed. 
Lower panel: The same analysis at LHC-7 with
$\sin \theta_{BL}$  taking on the values {(0.01,0.02,0.03,0.05)} from the bottom curve to the top in that order. }
\end{figure*}
\begin{figure*}[t!]\centering
\includegraphics[height=8cm]{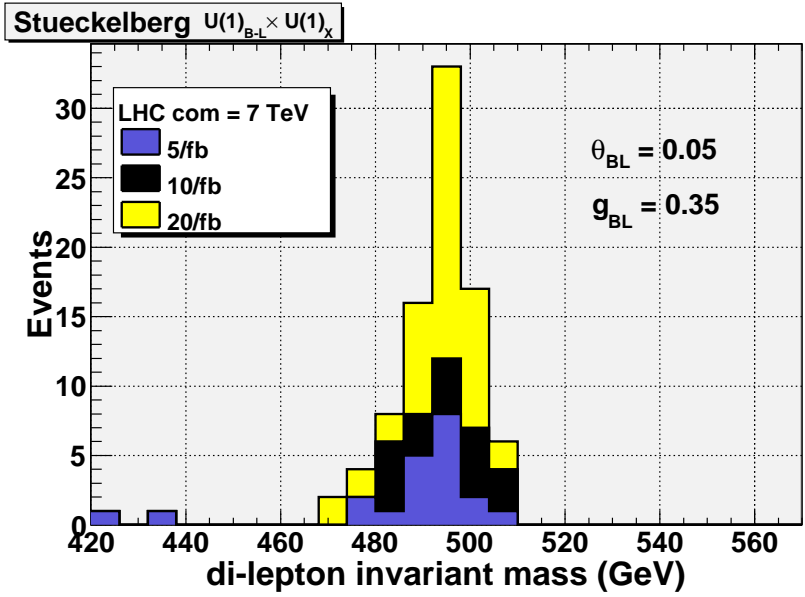}
\caption{
Exhibition of a 500~GeV  $Z'$ resonance in the Stueckelberg $U(1)_{B-L} \otimes U(1)_X$ model  at LHC-7 with
a  variable luminosity from 5fb$^{-1}$ to 20fb$^{-1}$ with a  $P_T$ cut on leptons of $P_T > 30$~GeV.
Currently   the LHC has analyzed $\sim \rm1fb^{-1}$ of luminosity.  For a $Z'$ resonance of 500~GeV with $\theta_{BL} = 0.05$ 
and $g_{BL} \sim g_Y$  the LHC would need about 5fb$^{-1}$ to begin to see any $Z'$ effect.
With  a very optimistic 20fb$^{-1}$, the $Z'$ signal will be strong and $Z'$ should be visible  with the 
mixings and masses of the size discussed.   }
\label{zp2}
\end{figure*}
In Table(\ref{Tab1}) we give the decay widths of the $Z'$ and $Z''$ bosons into leptons and into quarks.
The relative strength of the $Z'$ decay into quarks and leptons provides a distinctive signal for
this model. Thus, for example,  the ratio of the branching  ratios of $Z'$ into charged leptons 
vs into quarks (except into $t\bar t$) is given by 
$BR(Z'\to \ell^+\ell^-)/BR(Z'\to q\bar q) = 6/(5(1+ \alpha_s/\pi))$.
Further, in this model the decay width of the $Z'$ and $Z''$ are related by
\beqn
\frac{\Gamma (Z'\to \sum_i f_i \bar f_i)}{ \Gamma(Z'' \to \sum_i f_i\bar f_i)} = \tan^2\theta_{BL} \frac{M_{Z'}}{M_{Z''}}.
\label{ratiowidths}
\eeqn
Eq.(\ref{ratiowidths}) implies that for the $Z'$ mass in the sub TeV range, and the $Z''$ mass in the range
above  6~TeV, and $\tan\theta_{BL}\ll 1$ consistent with Eq.(\ref{c1}), the ratio of the decay widths
of $Z'$ vs of  $Z''$ can be vastly different, i.e., a decay width of $Z'$ in the MeV range vs the decay width
of $Z''$ in the hundreds of GeV range.  Thus while the $Z'$ will be a very narrow resonance,
the  $Z''$ will be a very broad resonance. 

It is also instructive to check the contribution of the new interactions to the muon anomalous moment which is
now measured very accurately~\cite{Bennett:2004pv} so that the current error in the determination is given by 
$\Delta (g_{\mu}-2)= 1.2\times 10^{-9}$. The contribution of the $Z'$ and of the $Z''$ bosons to the anomalous moment is
given by 
\beqn
\Delta (g_{\mu}-2) = \frac{g_{BL}^2 m_{\mu}^2} {24\pi^2} \left[ \frac{\sin^2\theta_{BL}}{M_{Z'}^2} 
+\frac{\cos^2\theta_{BL}}{M_{Z''}^2} \right].
\eeqn
Using the LEP constraint of Eq.(\ref{c2})
 one finds that the contributions of the new interactions is 
\beqn
\Delta (g_{\mu}-2) \leq \frac{ m_{\mu}^2}{12 \pi^2M_{BL}^2} 
\eeqn
and a substitution of $M_{BL}\sim 6$ TeV gives a rather small contribution, i.e., $\Delta (g_{\mu}-2) \leq O(1) \times 10^{-12}$. Remarkably in this case the LEP constraint of Eq.(\ref{c2}) is stronger than the constraint arising from
the very precise measurement of $g_{\mu}-2$.
\subsection{\it Production of Vector Resonances} 
The fact that the $Z'$ boson could have a low mass has important phenomenological implications.
From Table(\ref{Tab1}) we note that the decay width of the $Z'$ boson is proportional to $\sin^2\theta_{BL}$ 
and since $\sin\theta_{BL}$ is small, the decay width is relatively small, i.e., with the mass of the
$Z'$ in the sub TeV region, its decay width  would be in the  MeV range and thus
the \st $Z'$ is a very narrow resonance.
 A narrow resonance of this type
 should be testable in collider experiments much like   the hypercharge \st $Z'$
on which the D\O\ currently  has experimental bounds\cite{DzeroDY}. Further,  the decay of the
\st $Z'$ into leptonic  channels will be much more than in the hadronic channels
because the branching ratios  are proportional to $(B-L)^2$. Thus one can discriminate
a $ B-L $ \st $Z'$ boson by a study of its branching ratios.  Such a resonance  could be
produced in  the Drell-Yan process at the   LHC  and the Tevatron via 
\be 
pp (p \bar p) \to Z'\to \ell\bar \ell, q\bar q.
\ee 
In Fig.~(\ref{zprime1})  we show the predictions for the $Z'$  cross section times the branching ratio into $e^+ e^-$ 
in the $U(1)_{B-L} \otimes U(1)_X $ extension of the Standard Model.
Cross sections and event rates are calculated by implementing the couplings into PYTHIA and PGS~\cite{pythia,cway}.
The bottom panel shows the limits on the production cross section for  $\sigma(pp\to Z'\to e \bar e)$ at $\sqrt s = 7 ~\rm TeV$ with the recently released $\sim \rm 1~\rm fb^{-1}$ run~\cite{AtlasDY}. 
 For these curves we take $g_{BL}$ to be
have the same value as the hypercharge gauge coupling $g_{Y}$ and we let $\sin \theta_{BL}$ run from $0.01$ to $0.05$. The cross section for other values
of the product $g_{BL} \sin \theta_{BL}$ can be estimated by the scaling in the cross section which at the $Z'$ resonance scales like $g^2_{BL} \sin^2 \theta_{BL}$.
The top panel gives a similar analysis for the Tevatron using the D\O\ data with $\rm 5.4/fb$ of integrated luminosity~\cite{DzeroDY} .  
From the analysis of  Fig.(\ref{zprime1})  we observe that at present the Tevatron bound is about as strong as the
present LHC bound. However, the LHC will
surpass the Tevatron very soon.   Indeed, the $Z'$ produced in the model can exist with a much lower
mass~\cite{fln1,fln11,Salvioni:2009mt,Chanowitz:2011ew} than the $Z'$ models presently excluded by ATLAS~\cite{AtlasDY} and CMS~\cite{Chatrchyan:2011wq}.  
 In Fig.(\ref{zp2}) we  display the number of events as a function of the di-lepton invariant mass. 
 Here one finds that with an optimistic choice of an integrated luminosity of 20fb$^{-1}$ the number of 
 dileptonic events in excess of 30 in the peak mass bin  and should be visible. Thus a  $Z'$  mass of  500 GeV with a mixing angle
 $\theta_{BL}=0.05$ and $g_{BL}=g_Y$  is a promising candidate for discovery. 
\subsection{\it Production and Decay of the Scalars $\rho$ and $\rho'$}
In addition to the $Z'$ phenomenology there are other sectors where new phenomena can arise.
One of these relates to the scalar components $\rho_X$ and $\rho_{BL}$  
of $S+\bar S$ and of $S'+ \bar {S'}$ that remain in the bosonic sector after
 $Z'$ and $Z''$  gain mass  by the \st mechanism.  These fields mix  with the D-terms 
 so that one has the following set  in the Lagrangian
\beqn
 \rho_{BL} (M_1 D_{BL} + M_2' D_X) + \rho_X  
(M_1' D_{BL} + M_2 D_X).
\label{dterm}
\eeqn
  Elimination of the D-terms  gives the following mass matrix in  the $\rho_X$ and $\rho_{BL}$ basis
\beqn
 \label{neutrmass}
M^{2}_{\rm [spin~ 0]} = 
\left[
\begin{matrix}
 M_1^2  +  M^{\prime 2}_2  + m_{X}^2 & M_1M_1'+ M_2 M_2' \\
M_1M_1'+ M_2 M_2' & M^{\prime 2} _1+ M_2^2 + m_{BL}^2
\end{matrix}
\right] \ ,
\label{scalarmass}
\eeqn
where  we have also included the soft contributions to masses for $\rho_X$ and $\rho_{BL}$.
We note that the structure of the spin zero mass squared matrix given by Eq.(\ref{scalarmass}) is
different compared to the mass$^2 $ matrix given by Eq.(\ref{vectormass}). The reason for this
is that while the vector mass squared matrix arises directly from the  Stueckelberg term Eq.(\ref{mass}),
the mass squared matrix  of Eq.(\ref{scalarmass}) arises from the mixing given by Eq.(\ref{dterm}).
The mass matrix of Eq.(\ref{scalarmass})  gives two mass eigenstates $\rho$ and $\rho'$ with eigenvalues $M_{\rho}$ and 
$M_{\rho'}$.  
The mass parameters $M_1', M_2'$ can define the mixing and  when the mixing is small,
$M_{\rho}^2 \to M^2_1+ m_X^2$ and $M_{\rho'}^2\to M_2^2+ m_{BL}^2$.  With 
 $M_{\rho}$  in the sub TeV range  $M_{\rho'}$ may have a mass in the several TeV range. 
These mass eigenstates are admixtures of $\rho_X$ and $\rho_{BL}$ 
so that 
$
\rho_X= \cos \theta'_{BL} \rho + \sin\theta'_{BL} \rho'$
and $\rho_{BL}= -\sin\theta'_{BL} \rho + \cos\theta'_{BL} \rho'$.
For the case  when the soft terms are absent, the eigenvalues of the mass squared matrix
of Eq.(\ref{scalarmass}) is are identical  despite the very different looks of the matrices of 
Eq.(\ref{vectormass}) and Eq.(\ref{scalarmass}).
This can be seen by
the following unitary transformation 
\beqn
U^{\dagger} M^2_{\rm [spin~1]} U= M^2_{\rm {[spin~0]}}, \label{cool}
\eeqn
where the unitary matrix that connects the spin~1 and spin~0 matrix is given by
\beqn
U = \left(
\begin{matrix}
 \cos\xi & \sin\xi\\
 -\sin\xi & \cos\xi 
\end{matrix} \right) \ , ~~ \tan\xi= \frac{M_1'-M_2'}{M_1+ M_2}.
\label{UU}
\eeqn
This result shows that the  eigenvalues for the matrices $M^2_{[\rm spin~1]}$ and $M^2_{\rm[spin ~0]}$ are the same 
in the limit of vanishing soft masses for the scalars. 
 Now it is assumed that all the matter fields in the visible sector do not  carry any $U(1)_X$ quantum numbers,
i.e., $Q_X=0$ for quarks, leptons and the Higgs fields. Further, following the analysis of {Sec.(4)}, it is 
straightforward to establish that the quartic term $(\tilde\nu^{c\dagger} \tilde\nu^c)^2$ has a positive
co-efficient in the scalar potential. Thus once again since there are no couplings in the model to turn 
$M_{\tilde \nu^c}^2$  negative, there is no spontaneous violation of R-parity also in this extended model
while the 
$B-L$ gauge boson develops a mass via the \st mechanism.

From the discussion preceding Eq.(\ref{cool}) , it is clear that  the field  $\rho_X$ has no coupling with the visible 
sector while $\rho_{BL}$ has couplings of the form $g_{BL} M \rho_{BL} \bar{\tilde f}_i Q_{BL} \tilde f_i$.
One then has the following interactions of $\rho$ and $\rho'$ with sfermions  
\beqn 
{\mathcal L }_{\rho \tilde f^\dagger \tilde f} =- \sin\theta'_{BL} g_{BL} M_1 {\tilde f}_i^\dagger Q_{BL} \tilde f_i   \rho   
+ \cos\theta'_{BL} g_{BL} M_1 {\tilde f}_i^\dagger Q_{BL} \tilde f_i   \rho'.
\label{rho2}
\eeqn
Eq.(\ref{rho2}) allows the decay of the $\rho (\rho')$ via its couplings to the sfermions. 
If kinematically allowed $\rho (\rho')$ will decay into leptons + $E_T^{miss}$ or into jets + $E_T^{miss}$ 
where $E_T^{miss}$ contains at least two neutralinos  $\chi^0$  
(here ${\chi^0}$ is
the lightest neutralino (LSP) of the $U(1)_{B-L} \otimes U(1)_X$ combined sector - see sec.~(\ref{4})). However, an interesting situation arises
when the mass of $\rho (\rho')$ is smaller than $2 M_{\chi^0}$. In this case $\rho (\rho')$ cannot decay into the
final states with 2 LSPs and only the decays into the Standard Model particles are allowed. 
Such decays can occur  via loops and the final states will consist of $gg$,  $f_i\bar f_i$, 
$WW$, $ZZ$, $\gamma Z$, $\gamma\gamma$. There are many diagrams that contribute. 
 The dominant one
relevant to the model we study here with real scalars $\rho$ and  $\rho'$  
are the gluon fusion 
diagrams (see Fig.(3)).

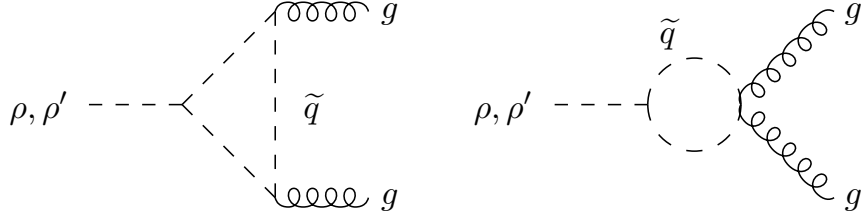
\begin{figure}[hbt]
\label{gluonX}
\begin{center}
\setlength{\unitlength}{1pt}
\scalebox{1.17}{
\begin{picture}(600,100)(80,0)

\Gluon(210,20)(240,20){-3}{4}
\Gluon(210,80)(240,80){3}{4}
\DashLine(210,80)(210,20){5}
\DashLine(210,20)(180,50){5}
\DashLine(180,50)(210,80){5}
\DashLine(150,50)(180,50){5}
\put(125,46){$\rho, \rho'$}
\put(220,46){$\widetilde{q}$}
\put(245,18){$g$}
\put(245,78){$g$}

\DashLine(300,50)(330,50){5}
\DashCArc(345,50)(15,0,180){5}
\DashCArc(345,50)(15,180,360){5}
\Gluon(360,50)(390,80){3}{5}
\Gluon(360,50)(390,20){-3}{5}
\put(275,46){$\rho, \rho'$}
\put(335,70){$\widetilde{q}$}
\put(395,18){$g$}
\put(395,78){$g$}

\end{picture} } \\
\setlength{\unitlength}{1pt}
\caption[ ]{Diagrams giving rise  to 
{the production of the \st scalars, $\rho,\rho'$}  at the lowest order.}
\end{center}
\end{figure}

From Eq.(\ref{rho2}) the interactions of $\rho$ and $\rho'$ to the mass diagonal squarks 
are given by the following interaction 
\beqn
{\mathcal L }_{(\rho, \rho')\tilde q ^{\dagger} \tilde q} & =& -g_{\rho} M_1
\cos(2\theta_{\tilde q_i}) \left(\tilde q^{\dagger}_{1i} \tilde q_{1i} \rho -\tilde q^{\dagger}_{2i} \tilde q_{2i} \rho
\right) -
g_{\rho} M_1
\sin(2\theta_{\tilde q_i}) \left(\tilde q^{\dagger}_{1i} \tilde q_{2i} \rho +\tilde q^{\dagger}_{2i} \tilde q_{1i} \rho
\right) \nonumber\\
&+& (\rho \to \rho', -\sin\theta_{BL}\to \cos\theta_{BL}).
\label{rhoint}
\eeqn
with the $B-L$ dependance encoded via     \be g_{\rho}= \frac{1}{3} g_{BL} \sin\theta_{BL}' ,\ee
and where $i$ runs over the squark flavors.
Now while the $\rho,\rho'$ vertices allow couplings with squark mass eigenstates, where the two states
couple to are either the same state or different states, the gluino only couples to squark
states, where both  states have the same mass. Thus in Eq.(\ref{rhoint}) only the interaction terms 
proportional to $\cos 2\theta_{\tilde q i}$ enter in the gluon fusion diagram. As such, the 
 decay width of the $\rho$ to gluons is given by 
\beqn
\Gamma(\rho \to gg) = \frac{g_{\rho}^2\alpha_s^2 M_{\rho}^3 M_1^2}{512\pi^3} 
\left|\sum_{a=1,2; i} (-1)^{1+a} \cos(2\theta_{\tilde q_i})
\frac{L_1(r_{ai})}{m_{\tilde q_{ai}}^2}\right|^2
\label{rhogg}
\eeqn
with $r_{ ai}= M_{\rho}^2/(4m_{\tilde q_{ai}}^2)$, 
  and $L_1(r)$ is a loop function defined by\cite{Djouadi:1998az}
\beqn L_1(r)= r^{-2} \left[ r- f(r)\right], ~~~~
	f(r) = \begin{cases}
	 \arcsin^2(\sqrt r)
		&r\leq 1 \\
	-\frac{1}{4} \left( \log \frac{1+\sqrt{1-r^{-1}}}{1-\sqrt{1-r^{-1}}} - i\pi\right)^2,
		&r>1~.
	\end{cases} 
	\label{L1}
\eeqn
As a consequence of the symmetry of gauge interactions one also has
\beqn
\Gamma(\rho'\to gg) = \frac{M_{\rho'}^3}{M_{\rho}^3} \cot^2\theta_{BL}'
\Gamma(\rho \to gg).
\eeqn
Further   the partonic production  cross section of $\rho$ is 
given by  
\beqn
\hat\sigma(gg\to \rho) = \frac{ g_{\rho}^2 \alpha_s^2 M_1^2}{256 \pi M_{\rho}^4} 
 \left | \sum_{a=1,2;i}
(-1)^{1+a}  \cos (2\theta_{\tilde q_{i}})  r_{ai} L_1(r_{ai}) 
\right |^2
 \delta(1- M_{\rho}^2/\hat s).
 \label{rh011}
\eeqn
The hadronic production cross section relevant to the search for $\rho$ at the LHC is $\sigma(pp\to \rho)$ 
and is given by a convolution with the parton distribution functions for the gluon, which at leading
order  in the narrow width approximation is given by
\beqn
\sigma(pp\to \rho)(s) =  
\tau_{\rho}\frac{d L_{gg}^{pp}}{d\tau_{\rho}} 
\hat \sigma(gg\to \rho).
\eeqn
Here $\sqrt s$ is the $pp$ center-of-mass energy,
$\tau_{\rho} = M_{\rho}^2/s$,
and ${dL_{gg}^{pp}}/{d\tau}$ is given by 
\beqn
 \frac{d{\it L}_{gg}^{pp}}{d\tau}= \int_{\tau}^1  \frac{dx}{x}  f_{g/p}(x,Q) f_{g/p}(\frac{\tau}{x}, Q), 
\eeqn
where $f_{g/p}$ is the parton distribution function for finding the gluon inside a proton with momentum
fraction $x$ at a factorization scale $Q$.  A numerical analysis shows that 
$ \sigma(pp\to \rho)$ can lie in the range O(1000)~fb  in the most optimal part of the parameter space
for producing the $\rho$.

  \begin{figure*}[t!]\centering
\includegraphics[scale=0.43,angle=0]{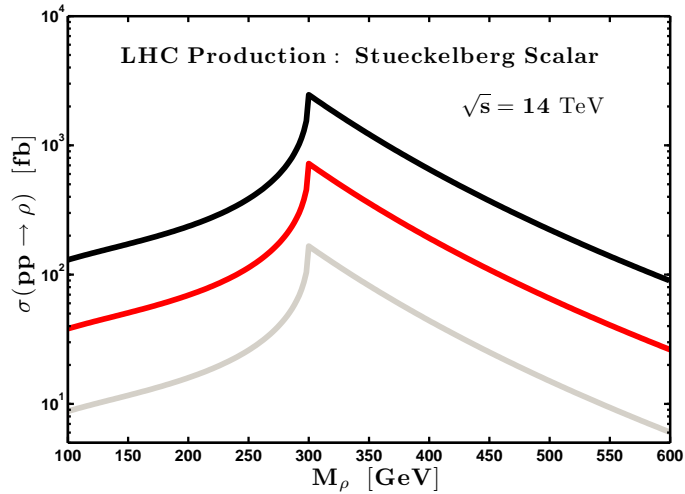}
\caption{A display of the production cross section   $\sigma(pp\to \rho)$ from gluon fusion as a function of the 
mass of the Stueckelberg scalar $\rho$ at the LHC  with $\sqrt s= 14$ TeV  for several combinations of $\theta'_{BL}$ and $g_{BL}$
for the case which maximizes
the production for the MSSM sector, $|\cos 2 \theta_{\tilde t}| =1$, i.e. $|A_t| \tan \beta = |\mu|$ .  From bottom to  top the curves have $(\theta'_{BL}, g_{BL}) =(\pi/6,0.65),(\pi/2, 0.65),(\pi/2,1.2)$,
where the top curve is close to the theoretical upper limit on the production. 
The kink appears at the point where $M_{\rho}/(2 m_{\tilde t_1})=1$
and the analysis has 
the other squarks and the gluino
much heavier than the lighter stop. }
\label{pprho}
\end{figure*}
 
The final decay modes of the $\rho$ can produce
 visible signatures at the LHC, and branching ratios
 will generally be different from the
 Standard Model Higgs $h_{\rm SM}$. Thus $h_{\rm SM}$ has both tree level decays into the final states
 $b\bar b, \tau\bar \tau, c\bar c$ as well as decays via loop diagrams into $gg, WW, ZZ, Z\gamma, \gamma \gamma$.
 For a Higgs boson mass of $100$ GeV, dominant decays modes are the tree level decay modes 
with $b\bar b$ decay being almost 80\%. Among the loop decays the dominant decay is $gg$ and
sub-dominant decays are $WW$ (off shell) and $\gamma\gamma$ at a Higgs mass of 100 GeV.  
 Now suppose the tree decays of the Higgs were suppressed, then the decay of the Higgs to
 $\gamma\gamma$ will have a branching ratio of $\sim 2.5\times 10^{-2}$.
 The decay of the $\rho$ parallels this case since there are no tree decays of the $\rho$. 
 In the analysis below we will use the above branching ratio to get an approximate estimate
 of $\gamma\gamma$ event for the $\rho$ decay. 
An analysis of $pp\to \rho$ at the LHC at 
$\sqrt s=14$ TeV is given in Fig.(\ref{pprho}). One finds that the cross section at  $M_{\rho}=100$ GeV 
for the maximal case with $(\theta_{BL}', g_{BL})=(\pi/2, 1.2)$ 
is $\sim 100$fb.  At 200 fb$^{-1}$ of integrated luminosity at the LHC at $\sqrt s=14$ TeV, one will have 
 $2\times 10^5$ $\rho$ events when $M_{\rho}=100$ GeV. Using $BR(\rho\to \gamma\gamma) = 2.5\times 10^{-2}$ 
 one finds  $\sim 5000$ $\gamma\gamma$ events before kinematic and efficiency cuts.  
 We note that the photons coming from the 
  $\gamma\gamma$ signal  will be monochromatic carrying roughly half the mass of the decaying particle. 
  Thus  the  $\gamma\gamma$   signal  arising from the decay of the $\rho$ would be distinguishable from the
 $\gamma\gamma$ signal from the Higgs decay if the masses  of the two are significantly separated.
 A $\rho$ mass of $100$ GeV would imply a $Z'$ mass of also 100 GeV assuming no soft terms in
 the $\rho$ sector. A $Z'$ mass of 100 GeV is consistent with the current data if either  
 the mixing angle $\theta_{BL}$ is small or the $Z'$ decays dominantly into the hidden sector
 (see Sec.(\ref{DD})). We note also that while the mass of the $Z'$ and the mass of $\rho$
 are the same in the absence of soft breaking terms for $\rho$,  the couplings of the 
  $Z'$ to fermions  and of $\rho$ to squarks can be of very different sizes. This is apparent from
  Eq.(\ref{UU}). Hence the possibility arises of being able to discover both the $\rho$ and the $Z'$.
  However it is also quite possible that only one resonance may be visible depending on the overall
  size of the Stueckelberg masses and the individual couplings of the two states.

The production cross section for $pp\to \rho, \rho' $ bears resemblance to the 
analysis of \cite{FileviezPerez:2011kd} and 
is closely related to canonical Higgs production~ (see e.g. \cite{Spira:1997dg,Djouadi:1998az})
but is restricted by the form of the couplings as given  in the $B-L$ \st extension.
We add that recently several models with scalars 
have been studied in the literature which can produce large production enhancements 
 relative to the SM higgs production (see e.g~\cite{Giudice:2000av,DeRujula:2010ys,Low:2011gn,Fox:2011qc}).
The production of $\rho$ does not receive enhancements of the size studied above, but nevertheless does
produce event rates that can be measured at the LHC-14 with larger luminosity as was detailed above.
 
  We note that very recently the LHC has put new constraints
on the allowed mass of the Standard Model Higgs Boson $h_{\rm SM}$. Preliminary analyses based on those reported
at  EPS 2011  and  at  Lepton-Photon 2011~\cite{EPSLP} imply that the SM  Higgs boson 
has a mass below $\sim$~145~GeV.   The  above result is compatible with the SUGRA models which typically
indicate a Higgs mass below $\sim 140~\rm GeV$. 
Because the production of $\rho$ relative to the $h_{\rm SM}$  differs markedly via their couplings,
as discussed above,   the  production of the 
two fields could be distinguished  with sufficient luminosity. This is possible if the $h_{\rm SM}$ 
resonance and the $\rho$ resonance are sufficiently separated in mass. In addition, 
because the production of $\rho$ is weaker than $h_{\rm SM}$, the golden channels
such as $ZZ,WW$ remain available  where  $h_{\rm SM}$
has been ruled out to have such a mass.  Searches for  $M_{\rho} \sim (200-500)\rm GeV$ 
will however have to wait for upgraded luminosity at the LHC.

\section{Neutral Dirac and Majorana Components of Dark Matter \label{4}}
\subsection{\it Majorana Dark Matter  }
The $U(1)_{B-L} \otimes U(1)_X $  \st extension of MSSM have new implications
for the nature of dark matter.
Specifically
 in  the neutralino sector  we have in addition to the MSSM neutralinos, extra gauginos and stinos,
 where the stino is the analogue of the higgsino.
 Thus from the gauge supermultiplets $X=(X_{\mu}, \lambda_X, D_X)$  and
 $C=(C_{\mu}, \lambda_C, D_C)$ 
 we can construct two gaugino states which we label as $\Lambda_X, \Lambda_{BL}$. Similarly from 
 the chiral multiplets $S+\bar S$ and $S'+\bar S'$ we can construct two higgsino states
 $\psi_S, \psi_{S'}$.  These four neutralino states in the 
 \st sector  have no mixings with the MSSM neutralinos. 
Thus the neutralino mass matrix   in the 
$U(1)_X\otimes U(1)_{B-L}$ extension of MSSM 
 has the form
 \beqn
 \cal{M}_{\rm neutralino} =
\left(
\begin{array}{c|c}
{\cal M}_{st} &  0_{4 \times 4} \\
 \hline
 0_{4 \times 4} &  {\cal M}_{\rm MSSM} \\
\end{array}
\right) 
\eeqn
Specifically the neutralino mass terms in the $U(1)_X\otimes U(1)_{B-L}$ sector are given by
\beqn
{\cal{L}}^{\rm mass} &=& - Z^T  {\cal{M}}_{st}  \  Z,   \\
Z^T &=& (\psi_S,   ~ \psi_{S'},  ~   \Lambda_{B-L}, ~   \Lambda_X),
\eeqn
where the $4\times 4$ sub-block of the ${U(1)_{B-L}\otimes U(1)_X}$ sector has the form
(omitting for simplicity the soft terms)
\beqn
 {\cal{M}}_{st}  = \left( \begin{array}{cc}
 0_{2\times 2}& m  \\
m^T & 0_{2\times 2} \\
 \end{array} \right)_{4\times 4},
 \label{e5}
 \  m = 
   \left( \begin{array}{cc}
 M_1 & M^{\prime }_2  \\
M^{\prime }_1 & M_2\\
 \end{array} \right)_{2\times 2}.
 \label{e5}
\eeqn
We can  diagonalize the neutralino mass matrix in the $U(1)_X\otimes U(1)_{B-L}$ sector
by  an orthogonal transformation $Z= O X$ so that 
\beqn
X^T O^T  {\cal{M}}_{st}  O X= diag(m_{\chi_5^0}, m_{\chi_6^0}, m_{\chi_7^0}, m_{\chi_8^0}) .
\eeqn
Now the generalization of the matter Lagrangian reads
\beqn
{\cal L}_{\rm matter} ~=~ 
 \bar \Phi_m e^{ 2g_{BL} Q_{BL} C +2g_{X} Q_{X} X } \Phi_m |_{\theta^2 \bar \theta^2}\ , 
\label{matter}
\eeqn and 
 gives a coupling of the type $\bar \Lambda_{B-L} f_L \tilde f_L^*$, 
where $(f_L, \tilde f_L)$ are a chiral fermion   and a chiral scalar, 
which leads to couplings of the \st sector neutralinos with matter of the form $\bar \chi_{k}^0 f_L \tilde f_L^*$
$(k=5-8)$.
Thus we  note that even though the neutralino mass matrix does not have a mixing between 
the MSSM and the \st sectors, the neutralino in the \st sector can decay into the least 
massive supersymmetric particle (LSP) which may lie in the MSSM sector. The way it occurs is as follows: The neutralinos
$\chi_k^0$ $(k=5-8)$ have fermion-sfermion interactions as  indicated above,
while the neutralinos in the MSSM also have similar 
type interactions.  If the mass $m_{\chi_k^0} > m_{\chi_1^0}$  we will have decays 
of the type $\chi_k^0 \to  \bar{\tilde f}_i f_i \to \bar f_i f_i \chi_1^0 + \cdots$. Thus there is only one stable
Majorana supersymmetric particle in the combined MSSM and \st system.
On the other hand if, for example,  $\chi_5^0$ is the lightest neutralino then the LSP will lie in the \st sector.
In this case the $\chi_{\rm st}^0 =\chi_5^0$ would
be a dark matter candidate {(the notation st denotes stueckelberg and does not imply preference of the stino
component over the gaugino component)}.  Its properties are expected
to be similar to those of the bino LSP of the MSSM.  
For the case of a thermal relic, the annihilation of $\chi_{\rm st}$ 
 will occur via the t-channel squark exchange  so that  (dropping the superscript 0 from here on)
$\chi_{\rm st} + \chi_{\rm st} \to  f_i \bar f_i,$ as wells 
$\chi_{\rm st}   + \tilde f_{\rm MSSM} \to  \rm SM ~SM', $
$\chi_{\rm st}   + \chi_{\rm MSSM} \to  \rm SM ~SM',$
where the last two cases indicate that the the coannihilations 
will generally occur ~\cite{darkforceA}, \cite{fucito},\cite{FLNN} (for a review see \cite{Feldman:1900zz}).
For the direct annihilations,  unlike the annihilation of MSSM neutralinos,  there are {\em no direct channel $Z$ or Higgs pole} exchange
diagrams and consequently final states such as $WW$, $ZZ$, $ZH$, $HH$  are absent at the tree level.    For the case of co-annihilations this is modified. {If $\rho$ is of low mass, as discussed in the previous section, the
stop should be relatively light to accommodate a signal of $\rho$. In this case
the relic density can be satisfied via  stop co-annihilations~\cite{Drees}. \\
 }

{
Next, we discuss the  the direct detection of $\chi_{\rm st}$. Specifically there are  {\em no t-channel Higgs or Z pole} exchange contributions to the direct detection 
rates for this case  at   the  tree level.
As pointed out in Ref.~\cite{Drees:1993bu,Hisano} it is important
to include contributions arising  in the spin independent scattering
cross section from the twist-2 operators
\begin{eqnarray}
f_p/m_p&\owns& \sum_{q=u,d,s} f_q f_{Tq}+ \sum_{q=u,d,s,c,b}
\frac{3}{4} \left(q(2)+\bar{q}(2)\right)g_q^{(1)}
  + \ldots
\end{eqnarray}
where  the additional terms are suppressed and $q(2)$, $\bar{q}(2)$  are matrix elements and are given in \cite{Hisano}.
 Specifically  $g_q^{(1)}$ is given by, in the limit of massless quarks
\begin{eqnarray}
g_q^{(1)}\simeq \frac{{M_{\chi_{ \rm st}}}}{(m_{\tilde{q}}^2-M_{\chi_{ \rm st}}^2)^2}\frac{a_q^2+b_q^2}{2}
\end{eqnarray}
where $a_q^2+b_q^2 = g^2_{BL} Q^2_{BL}=g^2_{BL} /9$.
In addition, there are terms
of size $\sum_{q=u,d,s} f_q f_{Tq}$ (where $f_q,f_{Tq}$ are given in \cite{Drees:1993bu,Chatto,Cor}).
Here terms in $f_q$ that are
proportional to $a_q^2+b_q^2$ are suppressed by a factor of 4 relative
to $g_q^{(1)}$~\cite{Hisano}. Terms in $f_q$ also contain $a^2_q-b^2_q  \propto g^2_{BL} Q^2_{BL} \sin 2\theta_{\tilde q}$
and are ultra suppressed by the smallness of the squark mixing angle.
For the case when the $M_{\chi_{\rm st}}$ is relatively close in mass to $m_{\tilde q}$, up to correction
in the light quark masses,  there is an enhancement in the
SI cross section\cite{Hisano}.  Utilizing this effect, for mass splitting of order 30-100 GeV,
one easily sees detectable size SI cross sections for squark masses that are in accord with
LHC limits (see Fig.(\ref{stsi})). At even smaller mass splittings, the models are constrained by XENON. 
We have verified  using  micromegas ~\cite{micro} that the small mass splitting between the LSP and the squarks
can lead to cross sections of the size we find. In this
case the relic density can be brought in accord with WMAP from co-annihilatons. In particular
the  squarks in the initial state annihilations play a large role in reducing the relic abundance.
There is also mixing 
 that derives from rotating between
the chiral fermion in the Stueckelberg multiplet.  We consider the optimal case where in
the mass diagonal basis, the lighter of the two mass eigenstates is the one which couples 
via the larger mixing.  Thus we have taken the mixing in the gaugino stino sector $\cos\theta_{\chi_{\rm st}} \to 1 $,
and have fixed $g_{BL} =0.65$ in Fig.(\ref{stsi}).
The result of a large scattering cross section  does
require an LSP above around (500-600)~GeV to be consistent
with the current limits from the LHC
 \cite{LHC1,LHC2,Akula1,Akula2,Buchmueller:2011ki}. 
}

\begin{figure*}[t!]\centering
\includegraphics[height=8cm]{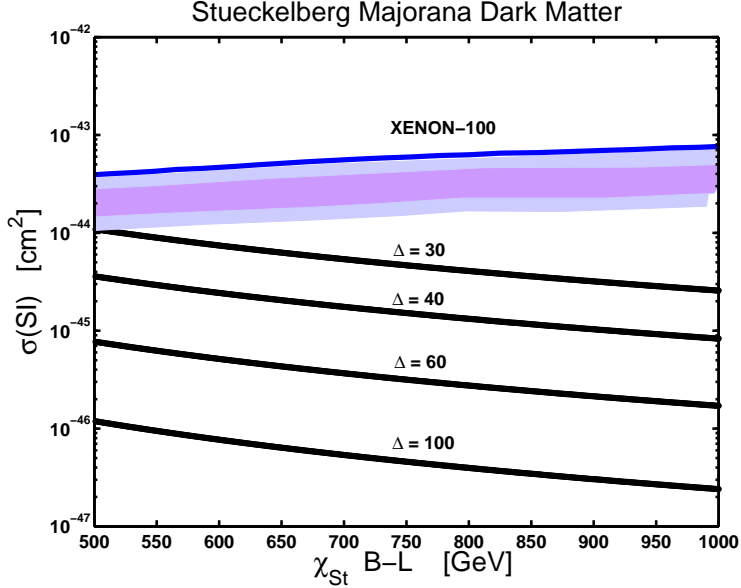}
\caption{
\label{stsi}
{Spin independent {$\chi_{\rm st}$} neutralino-proton cross section vs the Stueckelberg neutralino mass for the case when 
the Stueckelberg neutralino is the LSP. Exhibited is the spin independent cross section for several combinations 
of {$\Delta \simeq m_{\tilde q} -M_{\chi_{ \rm st}}$} (in units of GeV). The current limits from   XENON-100 are also exhibited.}}
\end{figure*}

\subsection{\it Dirac Dark Matter \label{DD}}
Additional matter fields in the form of Dirac fermions (and their supersymmetric counter parts, two
chiral scalars) can exist in the $U(1)_X$ sector which have only vectorial couplings to the gauge field 
$X^{\mu}$ and
a mass for the Dirac fermions can be generated via terms in the superpotential~\cite{Feldman:2010wy}. 
As seen already, after mixing of the $B-L$ gauge field $C^{\mu}$ with the field $X^{\mu}$, two mass eigenstates
$Z'$ and $Z''$ arise in the mass diagonal basis each of which have $B-L$ type couplings with 
the SM fields.  
In addition, the interaction of the dark sector Dirac field with the $Z', Z''$  is given by 
\be {\mathcal L}_{D}  =    \bar D \gamma^{\mu} (C_{Z' D} Z'_{\mu}+ C_{Z'' D}Z''_{\mu})D.  
\ee
The  interaction vertices with the Dirac  particle ($D$)  
with the visible  sector  quarks and leptons enter through the vector mixings  so that 
\beqn
C_{Z' D} &= &   g_X Q_X  \cos\theta_{BL},~~C_{Z'' D}  =
 g_X Q_X  \sin \theta_{BL}~. 
\eeqn
The dark sector Dirac field can constitute dark matter. It is stable and electrically neutral.
Since the model we consider has two components of dark matter, the total relic  density $\Omega h^2$
will be shared by the neutralino and
Dirac particles. In the analysis we  assume that the dark matter densities ${ \varrho}_D, { \varrho}_{\chi}$ for the two components in the galaxy 
are proportional to their respective  relic densities such that sum is the total cold dark matter (CDM) density 
\beqn
\frac{\varrho_D}{\varrho_{\chi}} \simeq  \frac{\Omega_D}{\Omega_{\chi}},~~~~~~~
{\Omega_{\rm CDM} h^2 }=  \underset{\rm (Majorana)}{\Omega_{\chi} h^2} + ~ \underset{\rm (Dirac)}{ \Omega_{D} h^2} .
\eeqn

The  annihilation cross section of $D \bar{D}$ into quarks and leptons via the 
$Z',Z^{\prime \prime}$ poles is given by 
 \beqn
\sigma_{D \bar D  \to f\bar f} &=& A_{D,f} |P_{Z'} - P_{Z''} |^{2}, 
\eeqn
where the poles and couplings enter as
\beqn
P_{V} &=&( {s-M_{V}^2+i\Gamma_{V} M_{V}})^{-1} , ~~  V= (Z', Z''), 
\\
A_{D,f} &=& \frac{g^2_{D,f} N_f}{4 8 \pi s}  (2 M^2_D+s )(2 m^2_f+s) \sqrt{\frac{4 m_f^2-s}{4
   M_D^2-s} } \tilde \Theta,  
\\ 
g_{D,f}   &=& 
  g_{BL} Q_{BL,f} g_X Q_X\sin 2\theta_{BL}, 
\eeqn
and where  $s = 4 M^2_D/(1-v^2/4)$,  $ \tilde \Theta  =\Theta( s-4 m_f^2)$, and  $N_f = (1,3)$  for  (leptons, quarks).
The relevant partial $Z', Z''$ decay {widths}  were given in Table~(1) . In addition the $Z',Z''$ can decay into the Dirac sector:
\beqn
\Gamma_{Z'\to D \bar D}=  \Theta \cdot \frac{M_{Z'} g^2_D}{12\pi}
\left(1+\frac{2M_{D}^2}{M^2_{Z^{\prime}}}\right)\left(1-\frac{4M_{D}^2}{M^2_{Z^{\prime}}}\right)^{1/2},
\label{aaa}
\eeqn
where ${\Theta= \Theta(M_{Z'} - 2 M_{D})}$  and $g_D ={g_XQ_X} \cos \theta_{BL}$. 
The partial decay width of the $Z^{\prime \prime}$ is obtained with  
$M_{Z^{\prime}} \to M_{Z^{\prime\prime}}$  and
$\cos\theta_{BL} \to \sin\theta_{BL}$ in Eq.(\ref{aaa}).
The relic density can be calculated by integration over the poles.
For the technique of integrating over  a pole see \cite{rd,rd1,rd2}.  
The relic density for the 2 components of dark matter can be calculated \cite{Feldman:2010wy}
 where for the Dirac component 
\beqn 
 \Omega_D h^2 = C_{D} J^{-1}_{D},~~~~
J_{D} = \int_0^{x_{F,D}}\sum_f  {\langle\sigma v\rangle}_{D \bar{D}  \to \bar f  f } ~dx, 
~~~~C_{D}  =  2\times \frac{1.07 \times 10^{9}~ {\rm GeV^{-1}}}{ \sqrt{g^*}M_{\rm pl}}.
\label{relicdiracabun}
\eeqn
In Fig.~(\ref{colorful}) we exhibit a satisfaction of the relic density within the WMAP constraint so that
\be R^{\rm St}_{\rm Dirac} \equiv \frac{M_D}{M_{Z'}}\simeq 1/2,\ee  where the black bands 
in Fig.~(\ref{colorful})  show
a presumed fraction of the the total relic abundance.
\begin{figure}[t]
\begin{center}
\includegraphics[width=14cm]{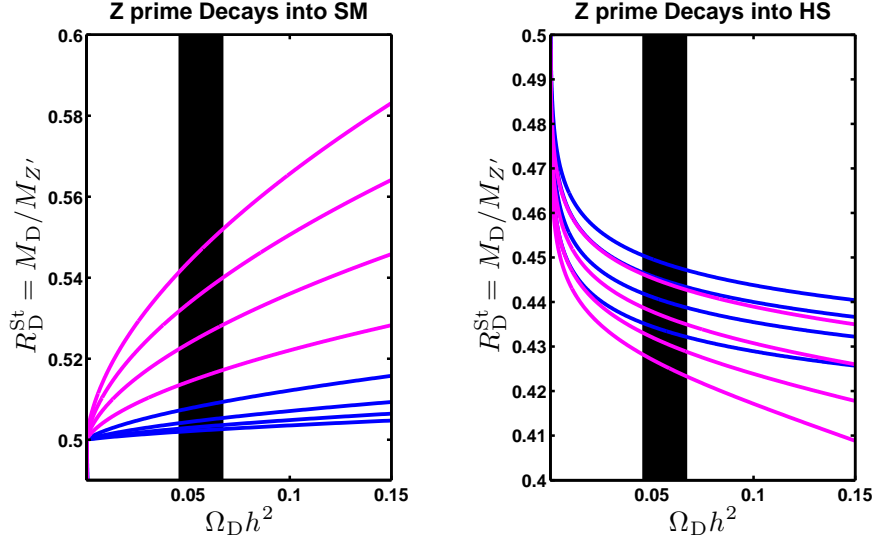}
\caption{An exhibition of the  relic density of the Dirac component of  dark matter for various values of 
  $R^{\rm St}_{\rm Dirac}$ which is the ratio the Dirac dark matter mass to  the Stueckelberg $Z'$ mass. The
  black bands represent about half the relic abundance.
  For the analysis we fix $g_{BL} = 0.35, g_X = 0.1, Q_X = 0.5$ .  The (blue/darker) curves have the $Z'$ mass running 
  in the range  200-500 GeV in steps of 100 GeV.  We note that
for fixed couplings,
as $M_{Z'}$ gets heavier the curves become more narrow.  
  The (magenta/lighter) curves correspond to $M_{Z'} = 250 ~\rm GeV$ with $\theta_{BL} = (0.02 -0.05)$.  
  Similarly, as $\theta_{BL}$ becomes progressively smaller for otherwise fixed couplings and fixed $Z'$ mass, the curves become more narrow.
The right panel is the case when the $Z'$ decays mostly into the hidden sector  Dirac fermions, i.e.,
it is the case  where  $Z' \to D \bar D $ is kinematically allowed and in this case 
 the dileptonic signals at the LHC will be depleted. 
  The left panel is the case where  $Z' \to D \bar D $ is kinematically disallowed and in this case 
 the $Z'$ will decay exclusively into the SM particles and thus 
 the dileptonic signal from the process $pp\to Z'\to l^+l^-$ will be visible. 
 \label{colorful}
}
\end{center}

\end{figure}

\begin{figure}[htb]\centering
\includegraphics[height=7cm]{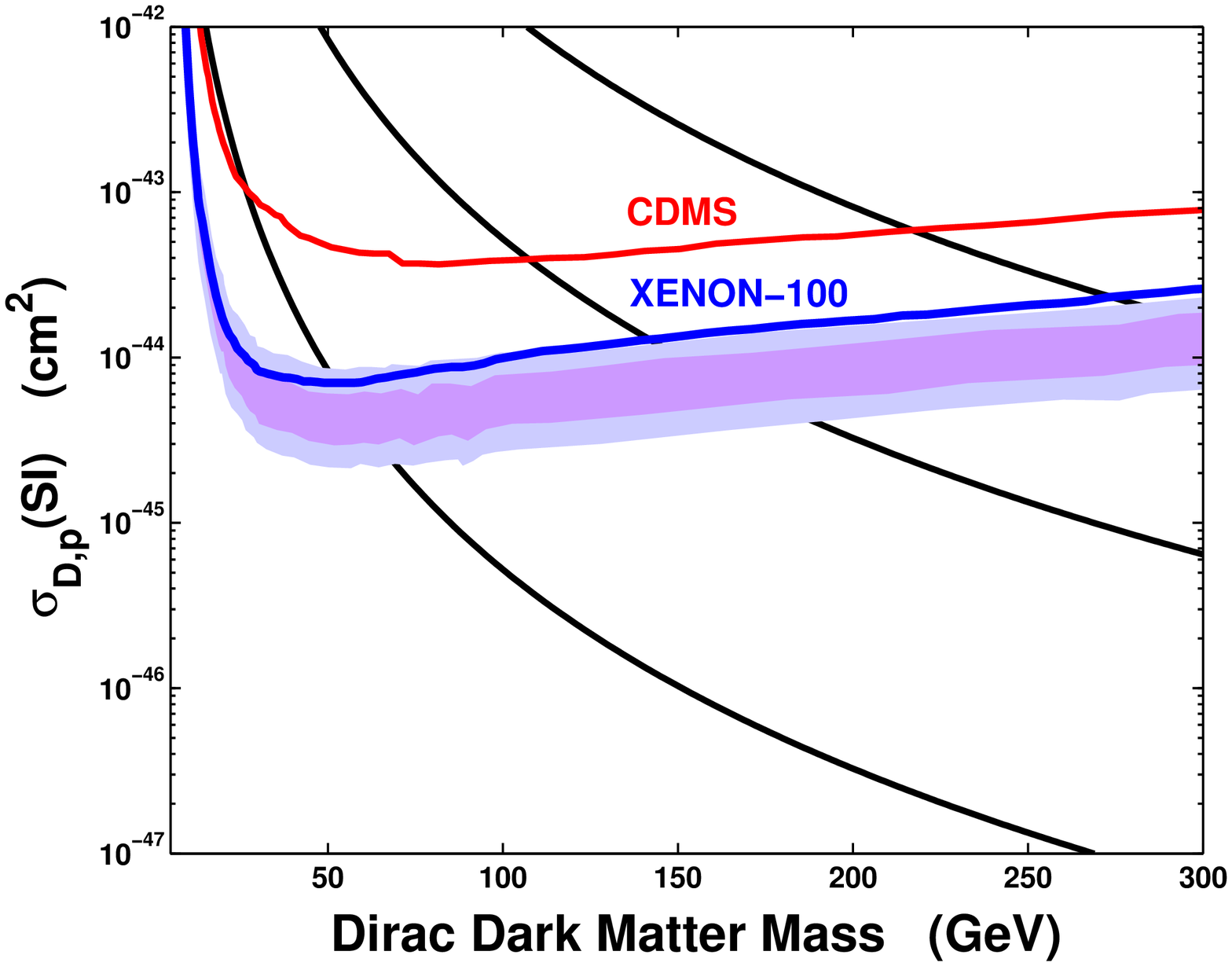}
\caption{ Illustrative curves. At any given point on this plot there exists a funnel where the relic density can 
be satisfied for perturbative size coupling via the relic density invariant $R^{\rm St}_{\rm Dirac}$. The particular
values of the parameters on thes curves are $(\theta_{BL}=.001, g_X =0.1, Q_X=1/2)$,  $(\theta_{BL}=0.01, g_X =0.5, Q_X=1)$, and $(\theta_{BL}=0.05, g_X =1/2, Q_X=1)$, 
where $g_{BL} = 0.35$ and $R^{\rm St}_{\rm Dirac} \sim 1/2$. }
\label{DDM}
\end{figure}
Now, unlike the cases studied previously  with the Stueckelberg mass growth,
 here the dark Dirac fermion does not carry a milli-charge and is electrically neutral.  
The reason the Dirac fermion  is neutral is because there is no mixing 
of the \st gauge field  with the hypercharge vector boson.
Because of its electrical neutrality and unlike a milli charged particle it cannot be stopped by
 the atmosphere or by dirt and  rock in the Earth before it reaches an underground detector.  
The effective Lagrangian describing the scattering of a  dark Dirac fermion from a quark,  in the limit of
 low momentum transfer, is given by  ${\mathcal L}_{\rm eff} = C^D_q \bar D \gamma^{\mu} D \bar q \gamma_{\mu} q$.
 The corresponding spin independent D-proton cross section is 
\beqn
\sigma_{D p}^{SI} =
 \frac{\mu^2_{D p}}{ \pi }  G^2
  \left(  \frac{1}{M^4_{Z'}} + \frac{1}{ M^4_{Z''} }  -\frac{2}{ M^2_{Z'} M^2_{Z''}   } \right ),
  \label{dddd}
\eeqn
where $G =  g_{BL}\sin\theta_{BL} g_X Q_X \cos\theta_{BL}$ and $\mu_{Dp}$ is the reduced mass. 

 \begin{table}[t!]
\begin{center}
\begin{tabular}{ccc}
 \hline
      $ M_{Z'}$  & $\sigma_{D p}^{SI} ~(\theta_{BL} = (0.03))$  & $\sigma_{D p}^{SI}~(\theta_{BL} = (0.06))$  \\
       $ \rm GeV$  &  $\rm cm^2$  &  $\rm cm^2$  \\\hline
     200  & 1.9$\times 10^{-44}$    &   7.5$ \times 10^{-44} $\\  
     300  & 3.7$\times 10^{-45}$    & 1.5$ \times 10^{-44} $   \\
     400  & 1.2$\times 10^{-45}$     & 4.7$ \times 10^{-45}$  \\
     500  & 4.8$ \times 10^{-46}$   & 1.9$ \times 10^{-45}$    \\
\end{tabular}
\caption{ Approximate values of the spin independent scattering cross section for the Dirac component of dark matter for sample models.
 The second and third columns have $\theta_{BL} = (0.03,0.06)$ respectively.  The first row is on the edge of the discovery limits from the both
XENON and the Tevatron data and is being probed by the LHC. For a given dark matter mass $M_{Z'} \sim 2 M_D$ in order to satisfy $(\Omega h^2)_{\rm WMAP}$.
Model parameters are otherwise fixed as in Figure(4). The middle column of this table corresponds to the blue/dark curves in Fig. (7), while the magenta/light
region is found to be constrained by the XENON data. Models consistent with the relic density constraint and the XENON constraint  are therefore favored if the relic density
is satisfied closer to the pole which is obtained for relatively smaller coupling and/or larger $M_{Z'}$.}
\end{center}
\end{table}

Interestingly, for mixing of the size considered in Fig.~(\ref{zprime1}),  ($\sin \theta_{BL} \in [0.01,0.05]$)  
and  for natural size couplings 
$g_X =g_{BL} =O( g_Y)  $ and  $Q_X = \pm 1$  one obtains a spin independent cross sections
which are  of the size  
\be 
\sigma_{D p}^{SI}  \sim 10^{-45\pm 1}  { \rm cm^2},  ~~~M_{Z'} \sim (200-300) ~\rm GeV.  
\ee
Since $M_D \gg m_p$, $\mu_{Dp}\sim m_p$,  $\sigma_{D p}^{SI}$ is essentially independent 
of $M_D$. 
However,  compatibility with the WMAP data for the 
thermal relic density, restricts the ratio $R_{Dirac}^{St}\simeq 1/2$.
Using this constraint the spin independent cross section $\sigma_{D p}^{SI}$ for the case  
$M^2_{Z''} \gg M^2_{Z'}$ is given by 
\beqn
\sigma_{D p}^{SI} \simeq
 \frac{\mu^2_{D p}}{ \pi }  G^2 \frac{1}{M^4_{Z'}}  \simeq   \frac{\mu^2_{D p}}{ 16 \pi }  G^2 \frac{1}{M^4_{D}} ~.
  \eeqn
which now has a very strong dependence on the Dirac mass. 
The numerical size of  $\sigma_{D p}^{SI}$ as a function of the Dirac mass is exhibited in Fig.(5), and the
analysis shows that the $\sigma_{D p}^{SI}$ predicted by the model is accessible in the XENON experiment.
In fact for given values of $g_{BL}, \theta_{BL}, g_X Q_X$ the current limits from XENON100 already 
put lower limits on the Dirac mass. We can also use the current upper limit on $\sigma^{SI}$ from the 
XENON100 experiment which gives $\sigma^{SI}=7\times 10^{-45}$ cm$^2$ for  a  WIMP mass of 50 GeV, 
to put a general constraint on $|G|/M_D^2$ so that
\beqn
|G|/M_D^2 \lesssim 3 \times 10^{-8}~~~~(M_D~\rm in~{\rm GeV}).
\eeqn
We note again that the preceding  analysis is very different from  the previous \st analyses where the 
Dirac fermion in the hidden sector  develops a milli charge. As already pointed out this arises in models
where one mixes
the \st gauge boson with the hypercharge gauge field. In this case the scattering of the Dirac
fermion from a  quark will have not only the $Z'$ pole in the t-channel but also a $Z$ boson pole and
a photon pole as well. 
In the present model the $Z$ and the photon pole are both absent. The Dirac dark matter candidate 
is electrically neutral.\\

As mentioned earlier, for $M_{Z'} \sim 2 M_D$, the relic density will always be satisfied for perturbative size couplings.
For $M_{Z'}< 2 M_D$ but close to $2 M_D$ the $Z'$ signal will manifest at colliders and the relic density can also be satisfied.
However, for the case $M_{Z'} > 2 M_D$, while the relic 
density can be satisfied, the $Z'$
 signal becomes suppressed due to the branching ratio into the hidden  sector overtaking the  branching ratio in the visible sector  in the presence of mass and kinetic mixings~\cite{diracdarkFLN}.
In addition, the Breit-Wigner enhancement of the annihilation of Dirac particles 
in the halo~\cite{Feldman} can be operative very close to the pole 
and the following three possibilities become simultaneous observables:
\begin{enumerate}
\item Observation of a very light  and narrow $Z'$ vector boson in the dilepton channel at the LHC (see also \cite{fln1}).
\item Observation of the flux of positrons via Satellite data  (PAMELA/FERMI)~\cite{PAMELA}  from the Breit-Wigner Enhancement in the dark matter annihilations in the galactic halo~\cite{Feldman} consistent with WMAP data~\cite{WMAP}.
\item Relic abundance of dark matter split between a neutralino and dark Dirac  (see also \cite{Feldman:2010wy}) .
\item Observational prospects for the corresponding Dark Dirac component  in  direct detections experiments such as XENON  (analyzed here for the neutral dark Dirac particle via the \st mechanism).
\end{enumerate}

Let us add, that just recently, the 730 kg days of the CRESST-II Dark Matter Search was
released~\cite{Angloher:2011uu}. Two preferred regions
are reported on, and one such region appears close to the CoGeNT preferred
region~\cite{Aalseth:2010vx}.  Very low mass
 neutralino dark matter with MSSM field content and cross sections
 of the size needed to explain the CoGeNT are not 
  consistent with the collider constraints~\cite{DFZLPNLowmass}. 
 This result has been confirmed by the LHC with
 its updated constraints on the SUSY Higgs sector~\cite{MV}, wherein large $\tan \beta$
 and  low mass SUSY Higgs of the size needed to explain the spin-independant
 scattering are further excluded. 
The  preferred  region reported by 
CRESST-II with heavier dark matter mass may be accommodated for a thermal relic with relic density 
satisfied via the Z-pole in the MSSM. Such could arise with non-universal gaugino masses
at the the high-scale (see~\cite{Akula1}) leading to WIMP masses
close to 45~GeV.  The far boundary of the CRESST-II  $2\sigma$ region terminating
close to 55~GeV  may also be
achieved with relic density satisfied via the Higgs pole~(see the analysis of \cite{Feldman:2011me}).
A dedicated analyses with the new constraints on the SUSY Higgs sector from the LHC~\cite{MV} would be 
needed to make a more definitive statement - however
the CRESST-II results at these potential dark mater masses do not correspond to 
reported event rates with CDMS or XENON~\cite{Ahmed:2009zw,Aprile:2011hi}. 
The extended model class we
discuss can produce spin independent cross sections
with larger cross sections than that of the MSSM via the Dirac component of Dark Matter (see Fig.(\ref{DDM})).

\section{Discriminating  \st  from Models with Spontaneous Breaking  \label{5}}
One may discriminate  between the \st mass growth for a $B-L$ gauge
boson in the models discussed here  and other models where the mass growth for
the $B-L$ gauge boson occurs
by spontaneous breaking. In the  above, we have already discussed the mass growth of a $B-L$ gauge
boson by the \st mechanism. For the case when the mass growth occurs via spontaneous breaking
 there are two possibilities: (i) spontaneous symmetry breaking of $U(1)_{B-L}$ occurs violating 
R-parity invariance, (ii) spontaneous symmetry breaking of $U(1)_{B-L}$ occurs without violating 
R-parity invariance.  We discuss these two cases below individually.\\
\subsection{Spontaneous Symmetry Breaking of $B-L$ and R-parity Violation}  
The simplest example of this is when we consider  the superpotential of 
Eq.(\ref{w1}). Let us assume that the potential of the $\tilde \nu^c$ field is such that it develops a 
VEV. In this case one will  have a spontaneous breaking of not only $B-L$ but also of
R-parity as indicated by the term $LH_u \langle \tilde  \nu^c \rangle$ in Eq.(\ref{w1}) after $\tilde \nu^c$ develops a VEV. 
In the mass diagonal basis it will lead to other R-parity violating terms, i.e., $LLe^c$
and $QLd^c$.  Here the LSP is no longer  stable and specifically the neutralino cannot
be a dark matter particle.  Further, since the neutralino is not stable, the signals of supersymmetry
for this case will be very different at hadron colliders. Specifically if the neutralino decays
inside the detector, there will be no missing energy signatures which are the typical
hallmarks of supersymmetry signatures with R-parity symmetry.  Further, for the case when there is 
a spontaneous breaking of  R-parity symmetry via the VEV growth of the right handed sneutrino, there
will be  D term contributions to the slepton squared masses proportional to $g_{BL}^2 \left< \tilde \nu^c \right>^2$\cite{FP2}. Such terms are
absent for the case when the mass growth for the $B-L$ gauge boson occurs preserving R-parity 
invariance as discussed  below. \\
\subsection{$B-L$ Models for R-parity Conservation}
We further consider now  the possibility that $B-L$ symmetry is broken but a residual R-parity symmetry 
still persists. This is indeed possible following the general line of  reasoning of 
\cite{Krauss:1988zc}  (see also \cite{Martin:1996kn}).
Thus consider  additional fields  in the theory such as a vector like 
multiplet which has the $SU(3)_C\otimes SU(2)_L\otimes U(1)_Y\otimes U(1)_{B-L}$ quantum numbers as
follows
\beqn
\Phi\sim (1, 1, 0, -Q_{BL}), ~~\bar \Phi \sim (1,1,0, Q_{BL}).
\eeqn
Let us suppose that one manufactures a potential so that VEV formation for the fields $\Phi$ and
$\bar \Phi$ occurs. In this case $B-L$ will be broken. However, as long as $3(B-L)$ is an even integer 
R-parity will be preserved. This means that the residual theory will have a $Z_2$ R-parity symmetry. 
Thus, for example, the VEV formation of  a scalar field 
 with $3(B-L)=\pm 2$ will violate $B-L$ but preserve  R-parity.
In the process of the  mass mass growth of the $B-L$ gauge boson, one combination of the imaginary parts of 
$\Phi^0$ and $\bar \Phi^0$ will be absorbed while there would three spin zero fields: 2 CP even 
 and one CP odd (the part orthogonal to the imaginary parts of $\Phi^0$ and $\bar\Phi^0$ which is absorbed)
 Higgs field. 
 In contrast for  the $U(1)_{B-L}\otimes U(1)_{X}$  model discussed here,
 one is left with only two additional scalars, $\rho_X, \rho_{BL}$, or $\rho, \rho'$,   which are both  CP even.
 Specifically there is no additional CP odd Higgs boson for the \st models. So this provides a discrimination
 between the two models. 
 
There are several  interesting and distinguishing features
 between the $U(1)_{B-L}\otimes U(1)_{X}$ model and the
 $U(1)_{B-L}$  model. This difference can be seen by comparing Eq.(\ref{c1}) vs
Eq.(\ref{c2}). Thus in Eq.(\ref{c1}) one finds that the mass growth of a $B-L$ gauge boson by
spontaneous breaking or by the \st mechanism 
would require the gauge boson to be very heavy. Thus for $g_{BL} \sim 1$, 
one will  typically have a mass of the $B-L$ gauge boson to be greater than $\sim~6~\rm TeV$~\cite{Carena:2004xs,Strumia}. 
In contrast, from Eqs.(\ref{c2}) we find that in the $U(1)_X\otimes U(1)_{B-L}$ model, there are two extra 
massive gauge bosons beyond what one has in the Standard Model. Thus the heavier  one, i.e., 
the $Z''$ gauge boson, is indeed several TeV in mass. However, the $Z'$ boson  we discuss 
can be much lighter, and can lie in the 
 few hundred GeV range. Thus the observation of a low lying $Z'$ with decay branching ratios
 characteristic of a $B-L$ gauge boson will be a clear indication of the \st   model involving mixing 
 of $U(1)_X$ and $U(1)_{B-L}$ discussed here.

\section{Conclusion \label{6}}
In this work we have proposed the \st  mechanism for  the mass growth of a  $B-L$ 
gauge boson.  {\textit{It was then shown that under the constraints of charge conservation and the absence of 
a Fayet-Iliopoulos D term, that  R-parity cannot be spontaneously broken  in the minimal model of radiative electroweak symmetry breaking}}.
The above is  in contrast to  models where the mass of the $B-L$ gauge boson is generated by the Higgs mechanism
 through the VEV formation for the field $\tilde \nu^c$ which breaks R-parity.
 
A comparison to the case where the $B-L$ symmetry is spontaneously broken but the R-parity symmetry is preserved
 was also given and its distinguishing features from the \st mass growth for the $B-L$ gauge boson 
 are uncovered. Further,  we analyzed a $U(1)_X\otimes U(1)_{B-L}$ \st extension of MSSM where a massive 
 $Z'$ boson with $B-L$ interactions can lie in the sub TeV region, i.e, $M_{Z'}< 1$~TeV.
 The observation of a $Z'$ in the sub TeV region with $B-L$ quantum numbers deduced
 via branching ratios into charged leptons will provide a test of the
 $U(1)_X\otimes U(1)_{B-L}$ \st extension discussed here. 
 
 Other tests of the proposed \st models  were also discussed.   This includes an analysis of the 
 production and decay of the \st spin 0 boson $\rho$ which has only loop decays into SM final 
 states via sfermion loops.
 An interesting decay of the $\rho$ is into $\gamma\gamma$  {was} analyzed and shown to have
  the possibility of observation at the LHC with $\sqrt s=14$ TeV.  
  
  With hidden sector Dirac fermions in the $U(1)_X\otimes U(1)_{B-L}$ \st extension,
   two component dark matter manifests,  with one component 
  being either the MSSM neutralino or the \st neutralino and the other component being a { {\it neutral} } Dirac
  fermion. An analysis of the relic density for the \st neutralino and  the  \st neutralino-proton  spin independent cross section were also discussed.  
 An analysis of the
  second dark matter component consisting of the Dirac fermion as dark matter was also given and it was shown that
 the current XENON100 data already puts constraints on the Dirac fermion mass and mixing angles.  The constraints
 from the XENON100 data and the LHC data on the couplings of the $Z'$ boson and dark Dirac fermion were shown to 
 be comparable, both of which limit the mixing of the $B-L$ and  dark sector. 
Thus the proposed model produces LHC and dark matter signals at mass scales
that are accessible to such experiments and will be tested further as the new data comes in.
\\

\noindent
{\bf Acknowledgments:}
P.F.P. would like to thank Northeastern University for hospitality in the beginning of this project.
The work of D.F.  is supported by DOE  DE-FG02-95ER40899 and by the Michigan Center for Theoretical Physics.
The work of P. F. P. is supported  by the James Arthur Fellowship at CCPP-New York University. 
The work of  P.N.  is  supported in part by NSF grant PHY-0757959 and PHY-0704067. 
D.F. would like to thank CERN Theory Group for their hospitality while this work was nearing completion.

\clearpage

\end{document}